\documentclass[twocolumn]{aastex631}
\usepackage[T1]{fontenc}
\DeclareRobustCommand{\VAN}[3]{#2}
\let\VANthebibliography\thebibliography
\def\thebibliography{\DeclareRobustCommand{\VAN}[3]{##3}\VANthebibliography}
\usepackage{graphicx}	\usepackage{amsmath}	\usepackage{amssymb}	\usepackage{threeparttable,tablefootnote} \usepackage{comment,hyperref}

\newcommand{\chisq}{\ensuremath{\chi^2}}

\newcommand{\mstar}{\ensuremath{M_\star}}
\newcommand{\msun}{\ensuremath{M_\odot}}
\newcommand{\mearth}{\ensuremath{M_\oplus}}
\newcommand{\xbar}{\ensuremath{\textbf{x}}}
\newcommand{\ybar}{\ensuremath{\textbf{y}}}

\newcommand{\hpp}{\ensuremath{\mathcal{H}_{PP}}}
\newcommand{\hppp}{\ensuremath{\mathcal{H}_{PPP}}}
\newcommand{\hpppp}{\ensuremath{\mathcal{H}_{PPPP}}}
\newcommand{\prog}[1]{\texttt{#1}}
\newcommand{\calceph}{\prog{CalcEph}}
\newcommand{\ttvfaster}{\prog{TTVFaster}}

\begin{document}
\title{Modeling the Solar System I: Characterization Limits from Analytic Transit Timing Variations}
\shorttitle{Solar System Characterization with TTVFaster}
\shortauthors{Lindor \& Agol}
\author[0000-0002-4061-3827]{Bethlee M. Lindor}
\affiliation{Department of Astronomy, University of Washington, Box 351580, Seattle, WA 98195, USA}
\affiliation{NSF Graduate Student Research Fellow}
\author[0000-0002-0802-9145]{Eric Agol}
\affiliation{Department of Astronomy, University of Washington, Box 351580, Seattle, WA 98195, USA}
\correspondingauthor{Bethlee M Lindor}
\email{blindor@uw.edu}

\begin{abstract}
Planetary systems with multiple transiting planets are beneficial for understanding planet occurrence rates and system architectures. Although we have yet to find a solar system analogue, future surveys may detect multiple terrestrial planets transiting a Sun-like star. In this work, we simulate transit timing observations of our system based on the actual orbital motions of Venus and the Earth+Moon (EM) — influenced by the other solar system objects — and retrieve the system's dynamical parameters for varying noise levels and observing durations. Using an approximate coplanar N-body model for transit-time variations, we consider test configurations with 2, 3, and 4 planets. 

For various observing baselines, we can robustly retrieve the masses and orbits of Venus and EM; detect Jupiter at high significance (for < 90-second timing error and baseline $\leq$ 15 yrs); and detect Mars at 5$\sigma$ confidence (with < 20-second timing error and baseline $\geq$ 27 yrs) using \ttvfaster. We also find that the 3-planet model is generally preferred, and provide equations to estimate the mass precision of Venus/Earth/Jupiter-analogues. The addition of Mars -- which is near a 2:1 mean-motion resonance with Earth -- improves our retrieval of Jupiter's parameters, suggesting that unseen terrestrials could interfere in the characterization of multi-planetary systems. Our findings are comparable to theoretical limits based upon stellar variability and \textit{may} eventually be possible.
\end{abstract}
\keywords{Astrobiology (74), Exoplanet detection methods (489), Solar system planets (1260), Transit timing variation method (1710)}
\section{Introduction}\label{sec:intro}
The current catalog of planetary systems has been populated primarily by the radial velocity (RV/Doppler shift) and primary transit discovery techniques. Although the \emph{Kepler Space Telescope} \citep{Borucki2010} was launched with the intent to discover habitable-zone \citep[HZ;][]{Kasting1993} Earth-sized planets transiting Sun-like (F/G/K) stars, it did not detect a true Earth-analogue planet. The \emph{Transiting Exoplanet Survey Satellite} \citep[TESS,][]{Ricker2015} is similarly unlikely to detect an Earth-Sun twin, since it was designed to detect hotter super-Earths around bright, nearby stars. 

Known exoplanetary systems have been found to contain planets with properties which differ dramatically from the solar system (SS). These systems display short-period gaseous planets \citep{Wright2012}, and a mix of compact super-Earths and mini-Neptunes \citep{Howard2013}.  Observational selection effects hinder the discovery of an SS-analogue: a cool gas giant with multiple inner terrestrial companions. However, upcoming RV programs show promise for detecting earth-mass planets in \added{a three-planet system that contains a 200$\mearth$ giant orbiting with} a 2953-day period for a 10 year observing baseline  \citep{Hall2018}. Nevertheless, the RV method provides only a lower limit on a planet's mass.

An opportunity for accurate characterization of an exoplanet system exists in a system that contains at least two transiting exoplanets. In a multi-transiting planet system (MTS), gravitationally-induced planet-planet interactions can significantly perturb the transit times \citep{Ragozzine2018}. \citet{Agol2005} and \citet{Holman2005} independently recognized that the resulting transit timing variations (TTVs) could be inverted to measure masses of perturbed, near-resonant terrestrials \citep[see also][]{MiraldaEscude2002,Schneider2004}. Moreover, the TTV analysis can reveal non-transiting planets that were previously undetected \citep{Ballard2011,Nesvorn2012}, and may allow us to probe relatively large orbital distances.

Photometric noise will affect the accuracy to which an observer can measure the planetary mid-transit times, and consequently, the accuracy to which they can characterize the system. Previous work finds that the mid-transit times of an Earth-like planet across a Sun-like star have a noise floor of 86 seconds \citep{Morris2020}\footnote{Adopted from Figure A1 of \citet{Morris2020}, which includes posterior distributions following a fit to the Sun's stellar variability in a single observing band.}. This noise floor is mostly due to $p$-mode oscillations affecting the disc-integrated stellar flux used in modeling the light-curves. However, simulations show that the precision of mid-transit times can be significantly improved by conducting multi-wavelength observations at very high signal-to-noise and decorrelating correlated noise across wavelength \citep{Gordon2020}. 

In anticipation of surveys which aim to characterize smaller masses/radii and longer orbital periods -- via transits (e.g., \textit{PLATO}, \citealt{Rauer2014}; \textit{Nautilus},  \citealt{Apai2019}) or direct-imaging (e.g., \textit{HabEx}, \citealt{Gaudi2020}; \textit{LUVOIR}, \citealt{Roberge2021}) -- and significantly improved precision, one can ask the question: what would the transits of SS planets reveal about the system. Answering this question would hopefully outline the survey design required to detect and accurately characterize a system that is analogous to ours. 

In this work, we perform blind transit timing analysis of a system with well-known parameters: our own. We simulate the orbital perturbations induced on two transiting terrestrial planets by treating our system as a proxy exoplanetary system. Of the terrestrials, Venus and Earth have the highest geometric probability of transiting the Sun, thanks to their sizes and proximity to the Sun \citep{Wells2017}. For the purposes of this current paper, we simulate SS orbits as viewed by an external observer, with one modification: we treat the Earth + Moon (EM) as one planet with the Earth-Moon dynamical barycenter (EMB) as its physical location. Our study of the Earth's transit times as perturbed by the Moon -- which imparts a TTV amplitude of 2.61 minutes on the Earth -- will be left for future investigation. For now, we assume the Earth and Moon are a single transiting body with a mass equal to the total mass of EM.

Given the remaining objects (i.e. Mercury, Mars, etc.) that are perturbing Venus and EM's orbits, we aim to find the required timing precision and observing schedule to address the following questions:
\begin{enumerate}
    \item Which non-transiting objects can we detect (e.g.\ Mars, Jupiter), and to what precision (and accuracy) can we characterize them?
    \item How precisely can we characterize the masses of the two transiting terrestrial planets (i.e. Venus and EM) with respect to the observing baseline and timing precision?
\end{enumerate}
In the following section, we describe our procedure for simulating (\mbox{\S\ \ref{sec:data}}) and fitting our suite of synthetic observations (\mbox{\S\ \ref{sec:modeling}}). In the first half of section \ref{sec:results}, we demonstrate our methodology for a fiducial case. We finish with a discussion of our overall results, and the impact of our findings. 
\section{Methods}\label{sec:methods}
To understand how a survey can affect the detection of SS-analogues, we simulated transits of Venus-like and EM-like terrestrials; by which we mean we adopted the physical and orbital parameters of these planets in our simulated observations. Using the \calceph\ package \citep{Gastineau2015:CALCEPH} -- which includes the gravitational influence of all substantial SS objects -- we compute the dynamical orbits for Venus and the EMB relative to the Sun via JPL ephemeris data. Given an SS object, a survey duration, and assuming white noise with a fixed standard deviation, we can create a simulated dataset of transit times. 

We then model transit times using \ttvfaster\ \citep[][hereafter AD16]{Agol2016}, and afterwards determine how accurately the masses and orbital parameters of the planets can be recovered -- including those whose properties can only be indirectly measured (i.e. Mars and Jupiter). This model assumes the planets are co-planar and edge-on in their orbits, which is appropriate as the RMS inclination of the eight solar system planets is less than 2$^\circ$\citep{Winn2015}.  

For each simulation, we perform a likelihood profile analysis to search for non-transiting objects, and globally fit the TTVs using the Levenberg-Marquardt algorithm \citep[LM;][]{Levenberg1944,Marquardt1963}. Next, we retrieve the masses and orbital elements of the planets by carrying out an affine-invariant Markov Chain Monte Carlo (MCMC) analysis \citep{Goodman2010}. Using both the likelihood profile and the MCMC sampling has benefits: the former checks for the modality of the probability distribution, and the latter characterizes the model parameter uncertainties. In the following sections, we describe our methods in more detail.
\subsection{Procedure for simulating data}\label{sec:data}
\subsubsection{Simulation specifications}\label{sec:simulations}
The JPL ephemerides have been generated by fitting numerically integrated orbits to both ground-based and space-based observations of SS objects. For this work, we adopt the DE440 version of the ephemerides to compute planetary position, velocity, and acceleration vectors at a given Julian Date \citep[JD;][]{Park2021}. Since a randomly-oriented, distant observer is unable to see all of the planets in the system transiting the Sun \citep{Brakensiek2016}, we assume that the observer lies along a line defined by the intersections of Venus and the EMB's orbital planes. Over numerous orbits, the location from which both Venus and EM could be seen transiting the Sun -- dubbed the transit zone -- would change very slightly due to planet-moon interactions, and planet-planet interactions \citep{Heller2016}. However, the individual transit zones for the terrestrials would be stable for thousands of years \citep{Wells2017}, and the timescales of our simulations are much shorter than secular timescales. 

To perform our simulations, we select an observing baseline for which we gather observations ($n_{\text{year}}$), starting at some time ($jd_0$). While the former varies, we fix the start time  to maintain consistency across our analyses\footnote{The DE440 JPL ephemerides are accurate over the epochs JD 2287184.5-2688976.5, and we chose $jd_0 =$ 2433282.5 JD}. We compute orbits from $jd_0$ until $jd_{\text{end}} = 365.25 \times n_{\text{year}} + jd_0$ with a time step of 22.5 days. In this paper, our observing duration ranges from 15 to 30 years, which encompass at least one full orbit of Jupiter ($\sim12$ years). This is a much longer observing duration compared to the periods of known exoplanets, but this length of time is needed to observe long-term TTV trends of the widely-separated SS planets. 

\begin{figure}
    \centering
    \includegraphics[width=\hsize]{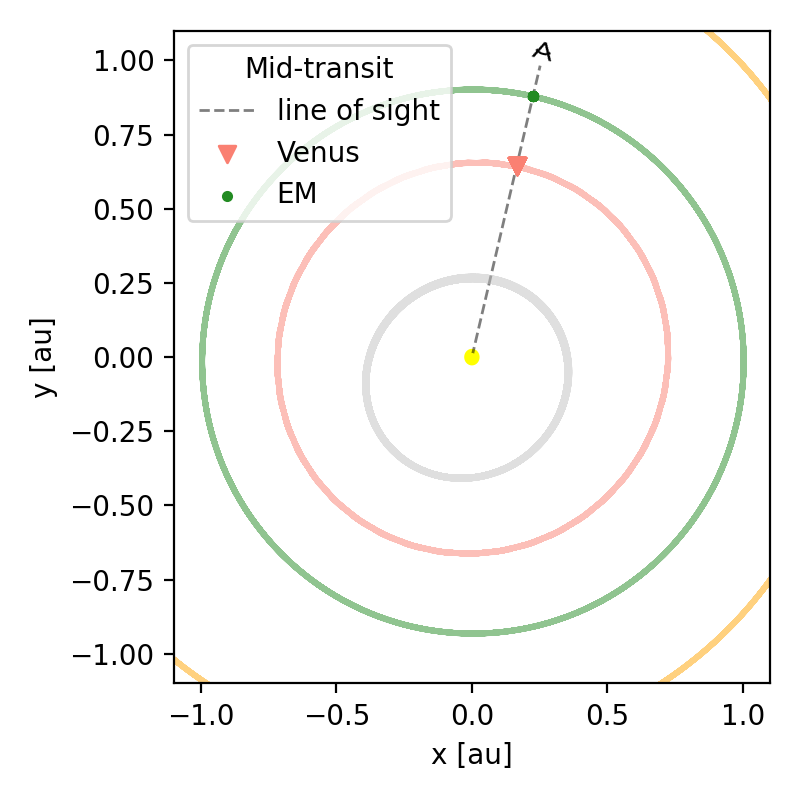}
    \caption{Top-down view of solar system orbits computed from JPL ephemerides over 15 yrs; the x-y axes represent the projected coordinates of the international celestial reference frame \citep[ICRF][]{Charlot2020} used by the JPL ephemeris. We indicate a projection of the direction near which the mid-transits of Venus (triangles) and EM (circles) can be seen.}
    \label{fig:orbits}
\end{figure}
\subsubsection{Mid-transit times}\label{sec:measurements}
During the orbit integration described above, we track the Cartesian coordinates and velocities to compute the sky-projected planet-star separation vector, $\textbf{r}_{\text{sky,}i}$, and its time derivatives $\textbf{v}_{\text{sky,}i}$ for the $i$th planet. Transit times are found by solving for roots of $\textbf{r}_{\text{sky,}i} \cdot \textbf{v}_{\text{sky,}i} = 0$ via the Newton-Raphson method. Figure \ref{fig:orbits} shows the locations where Venus and EM could be simultaneously observed transiting while orbiting the Sun over 15 years. The observer's line of sight is represented by the dashed line. 
\subsubsection{Injected timing noise}\label{sec:noise}
We assume that the noise can be mitigated below the stellar variability noise floor, and ran five simulations at a selected baseline, each with a different realization of white noise ($\sigma_\text{obs}$). For each mid-transit time, we draw the uncertainty from a Normal distribution with a standard deviation of $\sigma_\text{obs}$. After we add this uncertainty to the synthetic transit times, each dataset is given by $\textbf{y} = (\left\{t_{\text{obs},i}, \sigma_{\text{obs}};n^{\text{trans}}_{i}\right\};i=1,2)$; the subscripts $i=1$ and $i=2$ refer to Venus and EM, respectively. The total number of observations is  $N = \sum n^{\text{trans}}_{i}$, where $n^{\text{trans}}$ is the number of transits for each planet. 

Note that we use the same timing precision scale for both planets as they will differ only slightly in practice. Although Venus' transit is shallower relative to Earth, its faster orbital speed causes a sharper transit ingress and egress; this would yield timing precision similar to that of Earth (and thereby, EM). The timing errors -- which range from highly optimistic (10 seconds) to conservative (120 seconds) -- are summarized in Table~\ref{tab:sims}, along with our other observing parameters.   

\begin{table}
    \caption{Summary of simulation parameters.}
    \centering
    \begin{tabular}{ccc}
    \hline
    \hline
    Parameter & Values & Varied \\
    \hline
    $jd_0$ & 2433282.5 JD  & \\
    Baseline, $n_{\text{year}}$ & $[15, 16, ..., 30]$ & $\checkmark$\\
    $\sigma_{\text{obs}}$ [sec]&  $[10, 30, 60, 90, 120]$ & $\checkmark$\\
    $n_{1}^{trans}$ & $[24, 25, ..., 48]$& \\
    $n_{2}^{trans}$&$[15, 16, ..., 30]$& \\
    \hline
    \hline
    \end{tabular}
    \label{tab:sims}
\end{table}
\subsection{Fitting transit times}\label{sec:modeling}
Here we describe the transit timing analysis. We compute TTVs by comparing the observed transit times and those calculated assuming a linear ephemeris, which we review next.
\subsubsection{Transit ephemeris}\label{sec:calculations}
If the zeroth order Keplerian solution was correct for a planet, it would have a constant transit period ($P_i$) and mid-transit times given by:
\begin{equation}\label{equ:tcalc}
    t_{\text{calc},i} = t_{0,i} + P_i \times E ,
\end{equation}
where $t_{0,i}$ is the first transit time, and $E$ is the number of orbits elapsed after the initial transit time (i.e. epoch). For each planet, we fit this linear model to the noisy transit times, and derive an estimate for the mean ephemeris ($t_{0,i}$, $P_i$). Table~\ref{tab:example_sim} shows expected transit times of the linear ephemeris, and deviations from the mean period model for Venus and EM. In practice, transit-timing data often have missing or discontinuous observations, while in this paper we simulate contiguous transits based on the computed orbits. This observation was simulated for a 30-yr timescale with Gaussian uncertainties at the 30-s level. This dataset will be our fiducial case, which we will closely examine in sections \ref{sec:example}--\ref{sec:retrievals}. 
\begin{table*}[ht]
    \centering
    \caption{Transit timing data for Venus (planet b) and the Earth + Moon (planet c).}
    \begin{threeparttable}
    \begin{tabular}{cccccc}
    \hline
    \hline
    Planet & $n^{\text{trans}}$ & Noise-free Data & Mean transit ephemeris\tnote{a} & TTV & Injected noise\tnote{b}  \\
    &   & $t_{\text{obs}} - 2430000$ [JD] & $t_{\text{calc}} - 2430000$ [JD]& [min] & [min]  \\
    \hline
Venus & 0.0&3503.7644&3503.7657&-1.898&-0.06622\\
Venus & 1.0&3728.4658&3728.4665&-0.1012&0.9104\\
Venus & 2.0&3953.1687&3953.1672&2.229&0.1861\\
Venus & 3.0&4177.8674&4177.868&-0.9395&-0.07357\\
Venus & 4.0&4402.5666&4402.5688&-3.26&-0.174\\
Venus & 5.0&4627.2714&4627.2696&2.402&-0.2806\\
Venus & 6.0&4851.9724&4851.9703&3.537&0.5087\\
Venus & 7.0&5076.6718&5076.6711&1.399&0.3432\\
Venus & 8.0&5301.3698&5301.3719&-2.449&0.4775\\
Venus & 9.0&5526.0699&5526.0726&-4.609&-0.6018\\
    $\cdots$ & $\cdots$ & $\cdots$ & $\cdots$ & $\cdots$ &$\cdots$ \\
Earth+Moon& 0.0&3624.4054&3624.4024&3.818&-0.4276\\
Earth+Moon& 1.0&3989.6595&3989.6589&0.2987&-0.5574\\
Earth+Moon& 2.0&4354.9166&4354.9153&2.362&0.6001\\
Earth+Moon& 3.0&4720.1704&4720.1718&-2.334&-0.284\\
Earth+Moon& 4.0&5085.4292&5085.4282&0.9638&-0.4529\\
Earth+Moon& 5.0&5450.6864&5450.6847&2.417&-0.03565\\
$\cdots$ & $\cdots$ & $\cdots$ & $\cdots$ & $\cdots$ & $\cdots$\\
    \hline
    \hline
    \end{tabular}
    \begin{tablenotes}
        \item [a] $t_\text{calc}$ is found by linearly fitting the noisy simulated observed times.
        \item[b] We sampled a zero-mean Normal distribution with standard deviation of 1 and multiplied it by a scale factor of 30-seconds. 
        \item \textit{The entire table is available electronically in machine-readable format.}
    \end{tablenotes}
    \end{threeparttable}
    \label{tab:example_sim}
\end{table*}
\subsubsection{TTVFaster model}\label{sec:ttvfaster}
Next, we apply a retrieval to the synthetic transit times. The perturbations to the transit times are computed from formulae derived in AD16, which are added to the linear ephemeris. The \ttvfaster\ package implements these analytic formulae given a set of five parameters per planet: the planet-to-star mass ratio ($\mu_i = m_i/\mstar$), the orbital period ($P_i$) in days, the initial time of transit ($t_{0,i}$), the eccentricity ($e_i$), and the longitude of periastron ($\varpi_i$) as measured from line of sight. An MTS typically will have planets with nearly edge-on orbits \citep{He2019,He2020} therefore we can treat the problem in the co-planar approximation. Going forward, we let the longitude of periastron equal the argument of periastron ($\omega$). 

Suppose that we observe the transits of a body with mass $m_i$ orbiting a primary of mass \mstar. Following AD16, we define the unperturbed orbital frequency by $n_i^2 = G\mstar/a_i^3$, where $n_i = 2\pi/P_i$ is the mean motion. In the presence of a third mass $m_k$, the body $m_i$ will experience gravitational accelerations in addition to those caused by the primary. To describe the perturbations of the $k$th planet on the $i$th planet, we compute the inner-outer planet semi-major axis ratio as $\alpha_{ik}=(P_i/P_k)^{2/3}$ for $i<k$, and $\alpha_{ik}=(P_k/P_i)^{2/3}$ for $i>k$. The TTV solution is
\begin{equation}\label{equ:deltat}
        \delta t_{i,k} = \frac{P_i}{2\pi} \mu_{k} \sum_{j \geq 1} F_{i,k}^{j}(\alpha_{ik},\lambda_{i},\omega_i,e_i,\lambda_{k},\omega_k,e_k), 
\end{equation}
where $F_{i,k}^j$ is derived from the disturbing function -- and the sum is over an infinite series of sinusoidal functions. At mean-motion resonances, $pP_i^{-1} \approx (p+q)P_{k}^{-1}$ for integers $p$, and $q$ \deleted{for which} $F_{i,k}$ can become large. When truncated at a finite $j$, the sinusoidal functions approximate the timing variations caused by interactions between $m_i$ and $m_k$, with respect to the \mstar. We truncate the series at $j=5$ to yield a solution that is sufficiently accurate for our transit timing simulations.

We note that both longitudes ($\lambda_i$ and $\lambda_k$) in Equ.~\ref{equ:deltat} are evaluated at the calculated transit times for the $i$th planet in Equ.~\ref{equ:tcalc}. By summing over the perturbations induced in a system with $N_p$ planets, we can create planet $i$'s model transit times with the following: 
\begin{equation}\label{equ:tmod}
    t_{\text{mod},i} = t_{\text{calc},i} + \sum_{k\neq i}^{N_p} \delta t_{i,k} \text{.}
\end{equation}

\ttvfaster\ is as precise as N-body integration for a wide range of $\alpha$ values and eccentricities. Although the model tends to fail near mean-motion resonances, the low eccentricities ($\sim 0.01$) and  mass-ratios ($\sim 3 \times 10^{-6}$) of our two planets indicate better than 1\% precision for the observed planets\footnote{Figures 5 and 6 of AD16 show the fractional precision of the analytic formulae compared to N-body integration. The semi-major axis ratio $\alpha \approx 0.723$ in our case of Venus and EM}. Additionally, we expect \ttvfaster\ to be accurate for our solar system because each pair of planets in our model is external to first-order mean-motion resonance, and the RMS eccentricity of the eight planets is less than 0.09.

\subsubsection{Maximum likelihood}\label{sec:maxL}
Now, we derive our maximum likelihood function for the fit of Equ.~\ref{equ:tmod} to the transit times. In \ttvfaster, we parameterize in terms of the eccentricity vectors using the Poincaré variables: $h\equiv e\sin{\omega}$ and $k\equiv e\cos{\omega}$; the resulting parameter set for a system with $N_p$ planets sorted by period is given by $\xbar=(\left\{ \mu_i,P_i,t_{0,i}, k_i, h_i\right\};i=1,...,N_p)$.

Let us define the likelihood $\mathcal{L}$, as a function of the model parameters that represents the probability of measuring the observed data given the model, $p(\ybar|\xbar , \sigma_\text{sys})$ -- where $\sigma_{\text{sys}}$ is an unknown systematic uncertainty. Given the $j$th transit of the $i$th planet, we use the following log-likelihood function: 
\begin{equation}
\begin{split}
       \ln \mathcal{L}(\xbar,\sigma_\text{sys}) = & - \frac{1}{2} \sum_{i,j} \ln [2\pi ( \sigma_{ij}^2 + \sigma_{\text{sys}}^2)] \\
       &    -\frac{1}{2} \sum_{i,j}  \frac{(t_{\text{obs},ij} - t_{\text{mod},ij}(\xbar))^2}{\sigma_{ij}^2 + \sigma_{\text{sys}}^2}.
\end{split}
\end{equation}
Our model assumes that measured transit times deviate from the true transit times by independent and identically distributed random noise which has a zero-mean Gaussian distribution. Therefore, $\sigma_{ij} = \sigma_\text{obs}$, and we can rewrite $\ln \mathcal{L}$ in terms of the chi-square evaluated at $\sigma_{\text{sys}}^2=0$: 
\begin{equation}\label{equ:chi2}
    \chisq_0 \equiv \chisq (\sigma_{\text{sys}}=0) = \sum_{i,j} \Big( \frac{t_{\text{obs},ij} - t_{\text{mod},ij}}{\sigma_{\text{obs}}} \Big)^2.
\end{equation} The total log-likelihood for $N$ transit time observations is
\begin{equation}\label{equ:mclogL}
\begin{split}
    \ln \mathcal{L}(\xbar,\sigma_\text{sys}) = - \frac{N \ln{2\pi}}{2} - \frac{N \ln{(\sigma_{\text{obs}}^2 + \sigma_{\text{sys}}^2)}}{2} \\ - \frac{\chisq_0}{2} \frac{\sigma_{\text{obs}}^2 }{\sigma_{\text{obs}}^2 + \sigma_{\text{sys}}^2}.
    \end{split}
\end{equation}

We can remove the likelihood's dependency on $\sigma_\text{sys}^2$ via marginalization: 
\begin{equation*}\label{equ:logL}
\begin{split}
\mathcal{L}(\xbar)=\int_0^{\infty}  d(\sigma_{\text{sys}}^2) \mathcal{L}(\xbar,\sigma_\text{sys})=\\
      \int_0^{\infty}  d(\sigma_{\text{sys}}^2) \frac{(2\pi)^{-N/2}}{(\sigma_{\text{obs}}^2 + \sigma_{\text{sys}}^2)^{N/2}} & \exp{{\left(-\frac{\chisq_0}{2} \frac{\sigma_{\text{obs}}^2}{\sigma_{\text{obs}}^2 + \sigma_{\text{sys}}^2}\right)}}  \\
=(\chisq_0)^{1-N/2}  \sigma_{\text{obs}}^{2-N} \frac{2^{N/2 -1}}{(2\pi)^{N/2}} & \int_0^{\chisq_0 /2} x^\frac{N-2}{2} e^{-x} dx\\
     = (\chisq_0)^{1-N/2}  \sigma_{\text{obs}}^{2-N} \frac{2^{N/2 -1}}{(2\pi)^{N/2}} & \gamma(N/2 -1,\chisq_0/2).
\end{split}
\end{equation*}
As shown, the result is expressed in terms of the lower incomplete gamma function $\gamma(z,a)$  where $z= N/2 - 1$ and $a=\chisq_0 /2$ \citep[see][]{Andrews_Askey_Roy_1999}. $\ln \mathcal{L}$ becomes
\begin{equation}\label{equ:exact_logLmarg}
\begin{split}
\ln \mathcal{L}(\xbar) &= (1-N/2) \ln{\chisq_0}  + (2-N)\ln{\sigma_{\text{obs}}} - \ln{2}\\ 
& + \ln{\gamma(N/2 - 1,\chisq_0/2)} - N/2 \ln{\pi}.
\end{split}
\end{equation}
We use the property $\gamma(z,a)=\Gamma(z) P(z,a)$ when evaluating this equation, where $P(z,a)$ is the regularized lower incomplete gamma function. Since all terms aside from the first are nearly constant for each simulation, we can approximate the marginalized function with 
\begin{equation}\label{equ:approx_logLmarg}
\ln \mathcal{L}(\xbar) \approx  (1-N/2)\ln{\chisq_0} . 
\end{equation} 
When plotted, this distribution is more narrow than the exact solution but peaks at the same location. Therefore, the approximate expression is sufficient to find the maximum likelihood estimate (MLE).
\subsubsection{Test configurations and likelihood profile}\label{sec:tests}
Our pair-wise model allows us to decompose the transit times into the individual sources which perturb Venus and EM. With the peak amplitude ($A_{\text{TTV}}$) defined as the absolute maximum TTV value, we calculated the TTVs that these planets would experience due to the actual masses and orbital parameters of the SS objects. Based on these TTV amplitudes (see Table~\ref{tab:theory_effect}), we do not expect accurate retrievals for Mars and Jupiter unless the total timing uncertainty is $\leq 11.4$- and $34.8$-seconds, respectively. However, we proceed with our analysis as we are interested in how the timing error affects characterization. The TTV peak amplitudes for Venus and EM are $4.66$- and $5.97$- minutes, respectively. 
\begin{table}
    \caption{Theoretical TTV peak amplitudes based on their actual parameters. }
    \centering
    \begin{tabular}{llc}
        \hline
        \hline
        Perturber & Affected Object & $A_{TTV}$ [min] \\
        \hline
        Earth + Moon & Venus & 4.60 \\
        Mars & Venus & 0.19\\
        Jupiter & Venus & 0.58 \\
        Saturn & Venus & 0.10 \\
        Venus & Earth + Moon & 3.44\\
        Mars & Earth + Moon & 1.15\\
        Jupiter & Earth + Moon & 3.21 \\
        Saturn & Earth + Moon & 0.38 \\
        \hline
        \hline
    \end{tabular}
    \label{tab:theory_effect}
\end{table}

\begin{figure}
    \centering
     \includegraphics[width=.9\hsize]{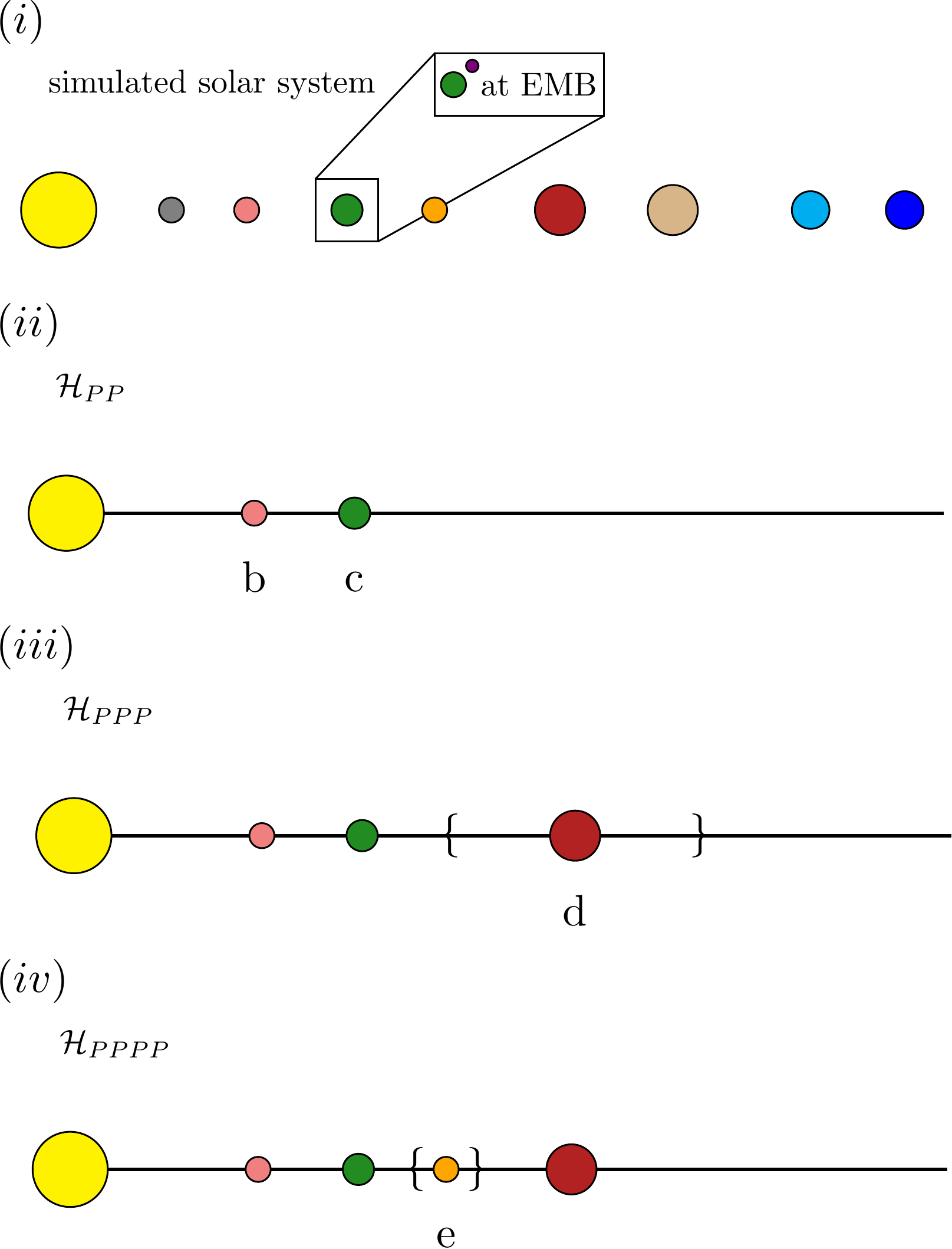}
    \caption{Schematic of the simulated multi-transit system (\textit{i}), along with the configurations considered in this work (\textit{ii}--\textit{iv}). We treat the Earth + Moon as planet c, with a location provided by the JPL ephemerides of the EMB. }
    \label{fig:schematic}
\end{figure}

For this paper, we test for several different model planet configurations, and compute the maximum likelihood ($\ln \mathcal{L}_{\text{max}}$) of each model given the transit times of Venus and EM -- henceforth dubbed planets b and c as if these were unknown exoplanets. An external observer would be unaware of the existence of non-transiting objects, therefore we begin by optimizing a preliminary 2-planet model. Figure~\ref{fig:schematic} illustrates the test configurations, which are all conditioned on observing Venus and EM as transiting planets. In this figure, a search for an additional planet is represented by a circle in a bracketed region. To detect unseen planets, we implemented a grid-based approach of maximizing the likelihood with an object added to the 2-planet model. This trial object is analogous to Jupiter (dubbed d) and Mars (dubbed e), depending on which configuration we are modeling. We do this progressively, which allows us to analyze the solar system as if it were an MTS, and to gauge where each hypothesis fails to describe the simulated transit times. 

\begin{figure}
    \includegraphics[width=\hsize]{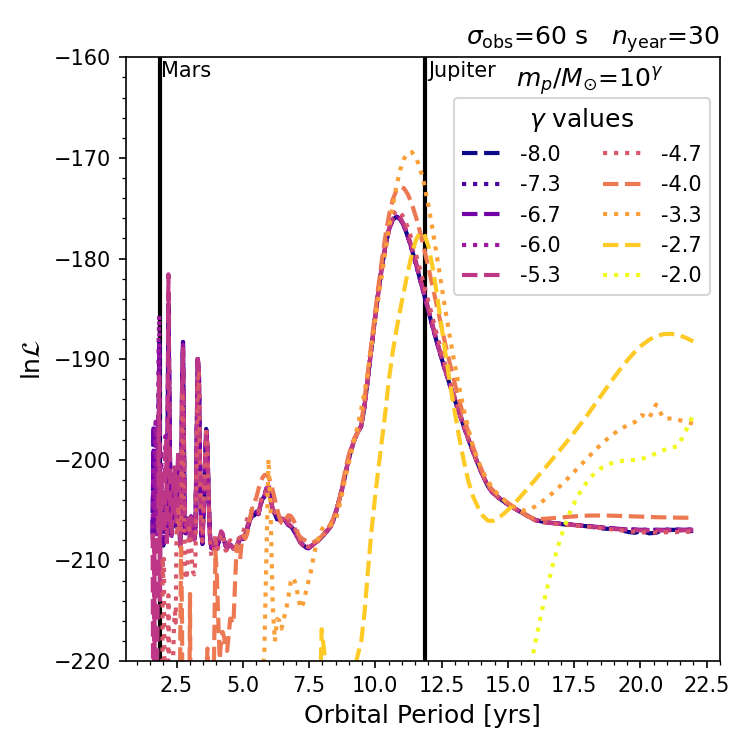}
    \caption{Likelihood as a function of mass-ratio and period of a third perturbing planet. Each curve corresponds to a value in ten grid points of the mass-ratio, where $\gamma = log_{10}(m_p/M_\odot)$. The log-likelihood was computed over 200 logarithmic steps in $P = [1.5,22]$ years, while optimizing over the orbital phase.}
    \label{fig:blind_search}
\end{figure}

We refer to our 3-planet hypothesis as $\mathcal{H}_{PPP}$. An initial blind test was performed with mass-ratios between $10^{-2}$ and $10^{-8}$, 200 periods in log-space ($P=[1.5,22]$ yrs), and 36 orbital phases for each period. First, we fix the orbital period, orbital phase, and mass-ratio at a grid value, and optimize all the remaining model parameters. We save the log-likelihoods computed during this process, and plot the likelihood profile as a function of $\mu$ and $P$ in Figure~\ref{fig:blind_search}. This figure shows that a third planet with $\mu=10^{-3.3}$ has the highest $\ln\mathcal{L}$ value when fitting 30 years of data with 60 seconds of injected Gaussian noise. We indicate the periods of Mars and Jupiter as vertical lines, which coincide with the regions of high log-likelihood. In practice, 60-s timing error would be too large to detect Mars, since it is larger than the peak TTV amplitudes that would induced on Venus (by 5 times) and Earth+Moon (by 9 seconds). 

We continue with a two-step process for calculating the MLE of the hypothesized 3- and 4-planet models, the latter of which is referred to as \hpppp. We then defined:
\begin{itemize}
    \item[1.] a coarse grid to locate the highest probability orbital phase and period, then
    \item[2.] a finer grid of 300 log-spaced points to refine our results. 
\end{itemize}
To the best-fit 2-planet model, we add a giant planet that has $\mu = 10^{-3}$, and broadly search over $P_d=[5,22]$ years. Then, we built fine grids around each peak: $P_d=[10.16,13.66]$, on average. We proceed to add a planet with a mass-ratio of $10^{-7}$ to our best-fit \hppp\ model. This \hpppp\ coarse grid is built with 200-400 points in log-space for $P_e=[1.5,5]$, while the fine grid ranges from 1.76 to 3.3 years on average. 

We summarize each hypothesis in Table~\ref{tab:models}. We adopt the results from the fine grid search for global optimization of each model. The subsequent fitted values are initial conditions for the MCMC sampling of our posterior probability.

\begin{table*}[!htbp]
    \centering
    \caption{Parameters used to find the maximum likelihood estimates.}
    \begin{threeparttable}
    \begin{tabular}{lc|cc|ccccc}
    \hline
    \hline
    \multicolumn{2}{c|}{}  &  \multicolumn{2}{c|}{Grid Search Values}  & \multicolumn{5}{c}{Starting Point}  \\
    Model & $N_p$ &  $\mu$ ($m_p/M_\odot$) & $P$ (years) &  $\mu$ & $P$& $t_0$ & $e\cos \omega$ & $e\sin \omega$\\
    \hline
    \hpp\ & 2 & -- & -- & $\mu_{b/c} = 3 \times 10^{-6}$ & \multicolumn{2}{c}{solutions to Eq.~\ref{equ:tcalc}} & 0.01 & 0.01 \\
    $\mathcal{H}_{PPP}$ & 3 & $[10^{-3},10^{-8}]$ & $[1.5,22]$ & -- & \tnote{\textdagger} & ... & 0.01 & 0.01 \\
    \hppp\ & 3 & -- & $P_d = [5,22]$ & $\mu_d = 10^{-3}$ & \tnote{\textdagger} & ... & 0.01 & 0.01\\
    \hpppp\ & 4 & -- &  $P_e=[1.5,5]$ & $\mu_e = 10^{-7}$ &  \tnote{\textdagger} & ... & 0.01 & 0.01\\
    \hline
    \hline
    \end{tabular}
    \begin{tablenotes}
    \item [\textdagger] We adopt the value corresponding to $\ln\mathcal{L}_\text{max}$ from the grid search (\mbox{\S\ \ref{sec:tests}}). 
    \item ``...'' denotes no starting point used in minimization. 
    \end{tablenotes}
 \end{threeparttable}
\label{tab:models}
\end{table*}
\subsection{Error analysis}\label{sec:errors}
Mathematically, Bayes' theorem states that $p(\xbar|\ybar) = p(\ybar|\xbar) p(\xbar)/p(\ybar)$ \citep{Bayes1763}. In terms of the likelihood, we can rewrite this equation to give the posterior probability distribution, $ p(\xbar,\sigma_\text{sys}) \propto \Pi(\xbar) \times \mathcal{L}(\xbar,\sigma_\text{sys})$ where $\Pi(\xbar) \equiv p(\xbar)$ is the prior function. While our approach to the likelihood in Eq.~\ref{equ:approx_logLmarg} does not account for the prior probability distribution, it provides an initial estimate with which to sample the posterior probability. 

Excluding the eccentricity vectors, we place a uniform prior on the planetary parameters. High-multiplicity planetary systems have lower eccentricities in order to remain stable \citep{He2020}. Therefore, we place an informative prior that each planet in our model has an eccentricity between 0 and $e_{\text{max}}=0.3$, with a gradual decrease in probability from $0.2$ to $e_{\text{max}}$. To ensure a posterior that has a uniform prior distribution in eccentricity, we sample the eccentricity vectors and weight their posterior samples by  ${((e \cos{\omega})^2 + (e \sin{\omega})^2)}^{-1/2} = 1/e$.

Using an affine-invariant Markov chain algorithm \citep{Goodman2010}, we sample our fitted parameter set and the systematic uncertainty, which is assumed to be the same for both planets. This parameter absorbs any inaccuracies in our model resulting in a posterior probability, $p(\xbar , \sigma_{\text{sys}})$. Given that high dimensional Markov chains require a large number of steps, we use $\approx 10^4$ steps with 75 random walkers. After the burn-in -- which we define as the first time the walkers cross the median parameter \citep{Knutson2009} -- the chain is examined to assess convergence and calculate the minimum effective sample size\footnote{Using \url{https://github.com/TuringLang/MCMCDiagnosticTools.jl/}}.
\section{Results}\label{sec:results}
In the following subsections, we compare how well the test configurations describe the TTVs. Then, we explore the distributions of the retrieved masses and dynamics (\mbox{\S\ \ref{sec:retrievals}}), and the effect of timing precision and observing baseline (\mbox{\S\ \ref{sec:effect}}).

\begin{figure*}
     \includegraphics[width=\textwidth]{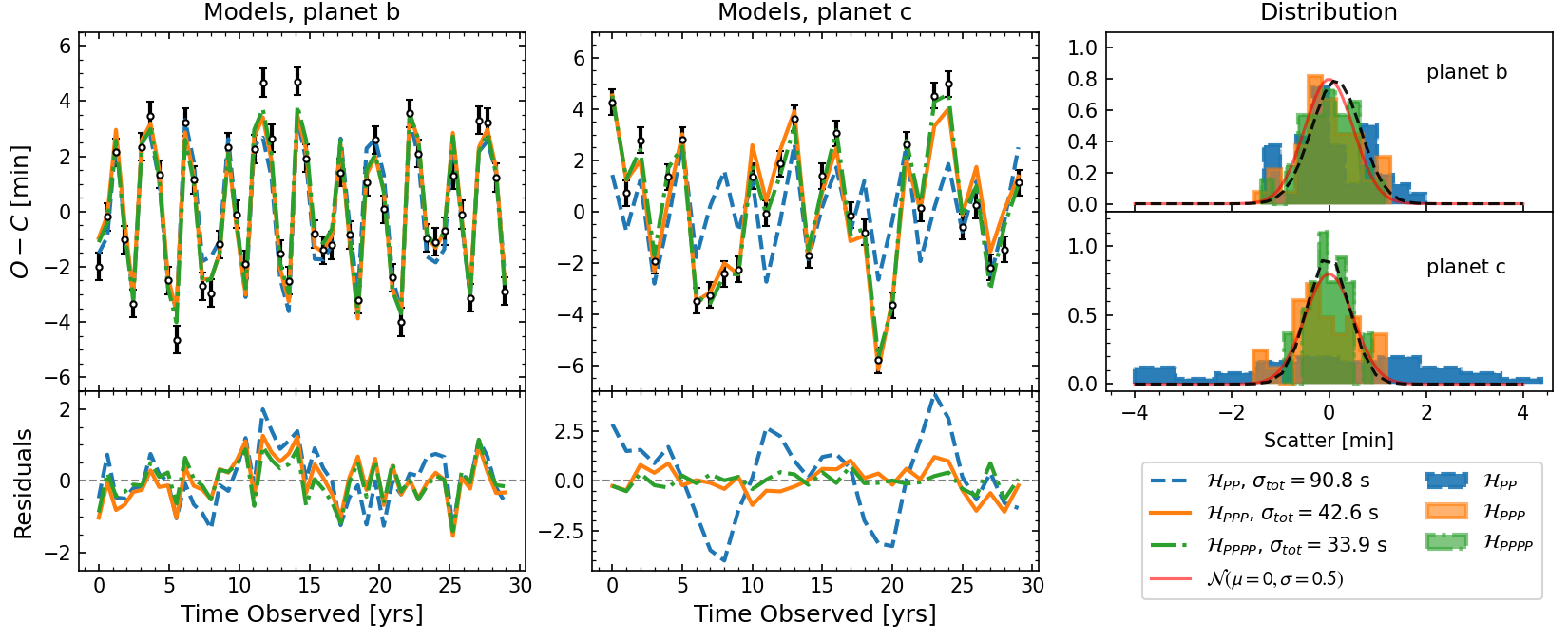}
    \caption{Model comparison for simulated observations of Venus (planet b) and Earth + Moon (planet c): \hpp\ in dashed blue, \hppp\ in solid orange, \hppp\ in dot-dashed green. \textit{Right:} histograms showing the distribution of scatter for each planet. The dashed black line corresponds to a Gaussian with the same mean and variance as the injected noise. The red curve is a Normal distribution with zero mean and a standard deviation of 0.5 minutes. }
    \label{fig:O_C}
\end{figure*}

\begin{figure*}[ht]
    \centering
\includegraphics[width=\linewidth]{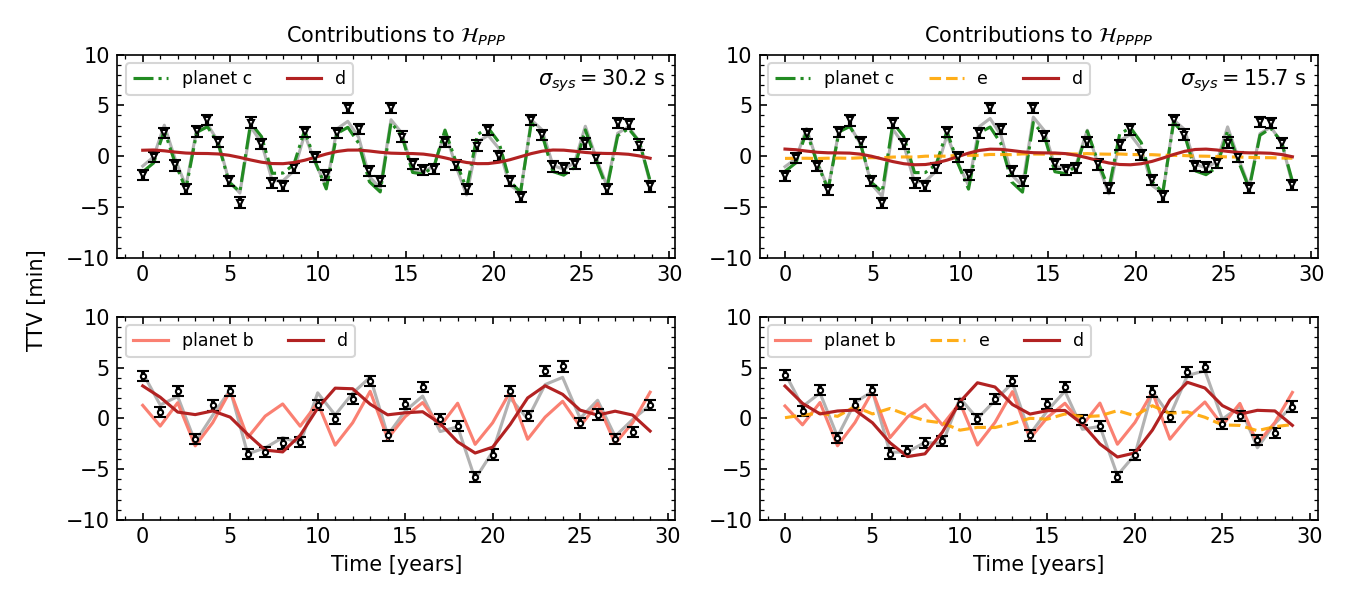}
    \caption{Simulated TTVs of 2 transiting terrestrial planets given in Table~\ref{tab:sims} -- Venus/b (top) and Earth + Moon/c (bottom) -- with a best-fit linear ephemeris removed, colored by source. \textit{Left panels}: The grey curve is the sum of all the perturbations, calculated from the median values of the posterior sampling of the best-fit 3-planet model. \textit{Right panels}: Same as the left, but here the solid grey line was created from the median posteriors in our best-fit 4-planet model.} 
    \label{fig:p3ttvs}
\end{figure*}

\begin{figure}
    \centering
    \includegraphics[width=\hsize]{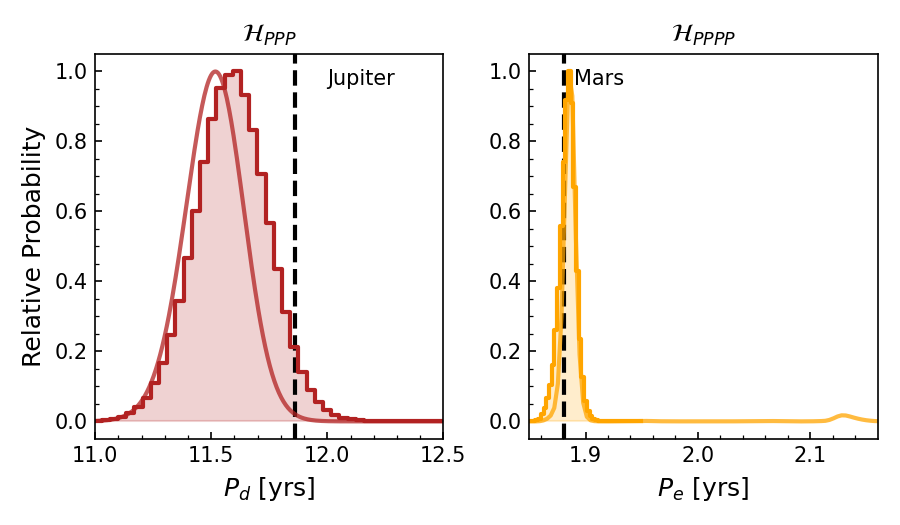}
    \caption{Period likelihood profiles for planets added in the \hppp\ (dark red curve) and \hpppp\ (orange curve) models, along with the histograms of the posteriors, for our fiducial case. The vertical lines indicate the actual periods of and Jupiter (left) and Mars (right).} 
    \label{fig:period_search}
\end{figure}
\subsection{Example simulations and fits}\label{sec:example}
We have selected the 80 transit epochs of Venus and EM as our fiducial case. This dataset is provided in Table~\ref{tab:30s30yrs}  with the $O-C$ residuals shown in Fig.~\ref{fig:O_C}. Overall, the two planet model has very large residuals -- with standard deviations of 0.78 and 2.1 minutes for planets b and c, respectively. The fact that the model with only 2 planets is insufficient for describing the observed transit times over long time spans suggests that there must be additional planets. Indeed, the best-fit models with additional planets reduces the scatter to 0.61 and 0.7 minutes (for $\mathcal{H}_{PPP}$), and 0.55 and 0.41 minutes (for $\mathcal{H}_{PPPP}$).

In Fig.~\ref{fig:O_C}'s right panel, we show a histogram of the model residuals compared to the injected noise (which we fit to a Gaussian). For Venus (i.e. planet b), the model residuals are all consistent with the injected noise and a Normal distribution. In contrast, planet c exhibits a wide spread of \hpp\ residuals and its scatter becomes 3 times smaller in the \hppp\ model, due to the stronger effect of Jupiter on EM (see Table.~\ref{tab:theory_effect}).

We plot each perturbation source for the best-fit 3- and 4-planet configurations in Fig.~\ref{fig:p3ttvs}, demonstrating the long term periodic TTVs from d and e with different colors and line styles. The grey lines are the total model TTVs computed from the best-fit median posteriors. Although the pairwise TTVs of planets b and c are anti-correlated on a short timescale (i.e. they dynamically interact), ignoring the low amplitude contribution of a Mars-like planet nearly doubles the systematic uncertainty from 15.7 to 30.2 seconds. When added in quadrature with $\sigma_\text{obs}=30$, the total uncertainties for each model are 90.8, 42.6, and 33.9 seconds, respectively.
\subsection{The search for additional planets}\label{sec:ex_search}
We fit the model (Eqn.~\ref{equ:tmod}) to our simulated data by optimizing the likelihood estimate (Eqn.~\ref{equ:approx_logLmarg}). Figure~\ref{fig:period_search} shows the estimated relative probability of planets e and d as a function of orbital period for each hypothesis. The relative probability is equal to $\exp{\left(\ln \mathcal{L} - \ln \mathcal{L}_{\text{max}}\right)}$ -- the maximum value of $\ln{\mathcal{L}}$ normalized to one. 

We summarize our sampling results for the fiducial case in Table~\ref{tab:30s30yrs}. For each model, we report the median posterior parameters and a 68.3\% quantile interval from the Markov chain analysis. 
We compute the Bayesian Inference Criterion \citep[BIC;][]{Schwarz1978,Kass&Raftery1995} with the \chisq\ that corresponds to the maximum value of Eqn.~\ref{equ:approx_logLmarg}. The BIC estimates the Bayesian evidence for a given model by assuming that the posterior obeys the Laplace approximation. This assumption is not valid for all the posteriors (see the eccentricity vectors in Figure~\ref{fig:1d_hist}), but we can still use BIC to penalize models with additional free parameters. 
 With respect to the maximum likelihood, 
\begin{equation}
   \text{BIC}=-2 \ln{[\mathcal{L}_{\text{max}}(\xbar)]} + k \ln{N}\text{,}
\end{equation}
where $k$ is the number of free parameters. 
\subsubsection{Two planets}\label{sec:p2_fit} 
The \hpp\ model has a $\chisq_0$ of 582.9. Although the systematic uncertainty is 85.8 seconds, the derived masses and eccentricities are consistent with those of Venus and EM: $m_b=0.778^{+0.161}_{-0.148}$ and $m_c=1.012^{+0.123}_{-0.107}$,  $e_b=0.019_{-0.002}^{+0.061}$ and $e_c=0.016_{-0.001}^{+0.051}$.
\subsubsection{Three planets}\label{sec:p3_fit}
We show the detection of a unique period for planet d in Fig.~\ref{fig:period_search}; also, left panel of Fig.~\ref{fig:multipeaks}. Orbiting every $4232.5^{+56.4}_{-57.0}$ days, this unseen planet  would have a mass of $m_d=266.5^{+56.1}_{-49.1} \mearth$. Although the error on planet d's mean ephemeris is quite large, this scenario is a decisive improvement over the 2-planet model; it reduces the BIC by 100, a 10$\sigma$ detection of Jupiter (Fig.~\ref{fig:deltaBIC}). Consequently, the masses and eccentricities of planets b and c are more accurate and well constrained: $m_b=0.836^{+0.074}_{-0.073}$, $e_b=0.011_{-0.001}^{+0.020}$; and $m_c=1.035^{+0.051}_{-0.049}$, $e_c=0.018_{-0.003}^{+0.013}$. 
    
Figure~\ref{fig:cornerplot_p3} displays a corner plot of the mass-ratios, periods, and eccentricity-vectors of this configuration. For the inner planets, the masses and periods follow Gaussian distributions, while the constraints on the eccentricity vectors are strongly correlated. We attribute this aspect to the eccentricity-eccentricity degeneracy described in \citet{Lithwick2012}. We also find that planet d (i.e. Jupiter) has a negatively skewed e-vector; and the  $e_d\cos{\omega_d}$-$\mu_d$ panel demonstrates a banana-shaped probability distribution function. This approximate degeneracy between masses and orbit shapes is known to impact measurement of planet masses due to observations with insufficient timing precision \citep[see][]{HaddenLithwick2017,Leleu2023}. 
\subsubsection{Four planets}\label{sec:p4_fit}
The BIC of the \hpppp\ model only differs from the \hppp\ model by 17, corresponding to a 4.1$\sigma$ detection of Mars. As such, we cannot confidently claim the discovery of this terrestrial for the fiducial example. Also, Fig.~\ref{fig:period_search} shows an -- albeit very small -- secondary peak in the likelihood profile. Upon inspection, the posterior trace for $P_e$ revealed that many walkers continue to sample low probability regions with large systematic uncertainties, despite agreeing closely with the likelihood profile near the maximum likelihood. 

The retrieved planet masses for this model are  $m_b=0.819_{-0.058}^{+0.061}$ and $m_c=1.033_{-0.040}^{+0.042}$, $m_d=319.3^{+58.9}_{-59.4}$, and $m_e=0.099^{+ 0.039}_{-0.028} \mearth$, in line with their true counterparts. However, the recovered eccentricity for planet e is inconsistent with that of Mars (see Tab.~\ref{tab:30s30yrs}). By examining the posterior distributions for planet e (Fig.~\ref{fig:cornerplot_p4}), we see that all of its components are distributed asymmetrically with long tails. Again, we note the mass-eccentricity degeneracies, which are more pronounced for planet e than d. 
\begin{figure*}[ht]
    \centering
    \includegraphics[width=\linewidth]{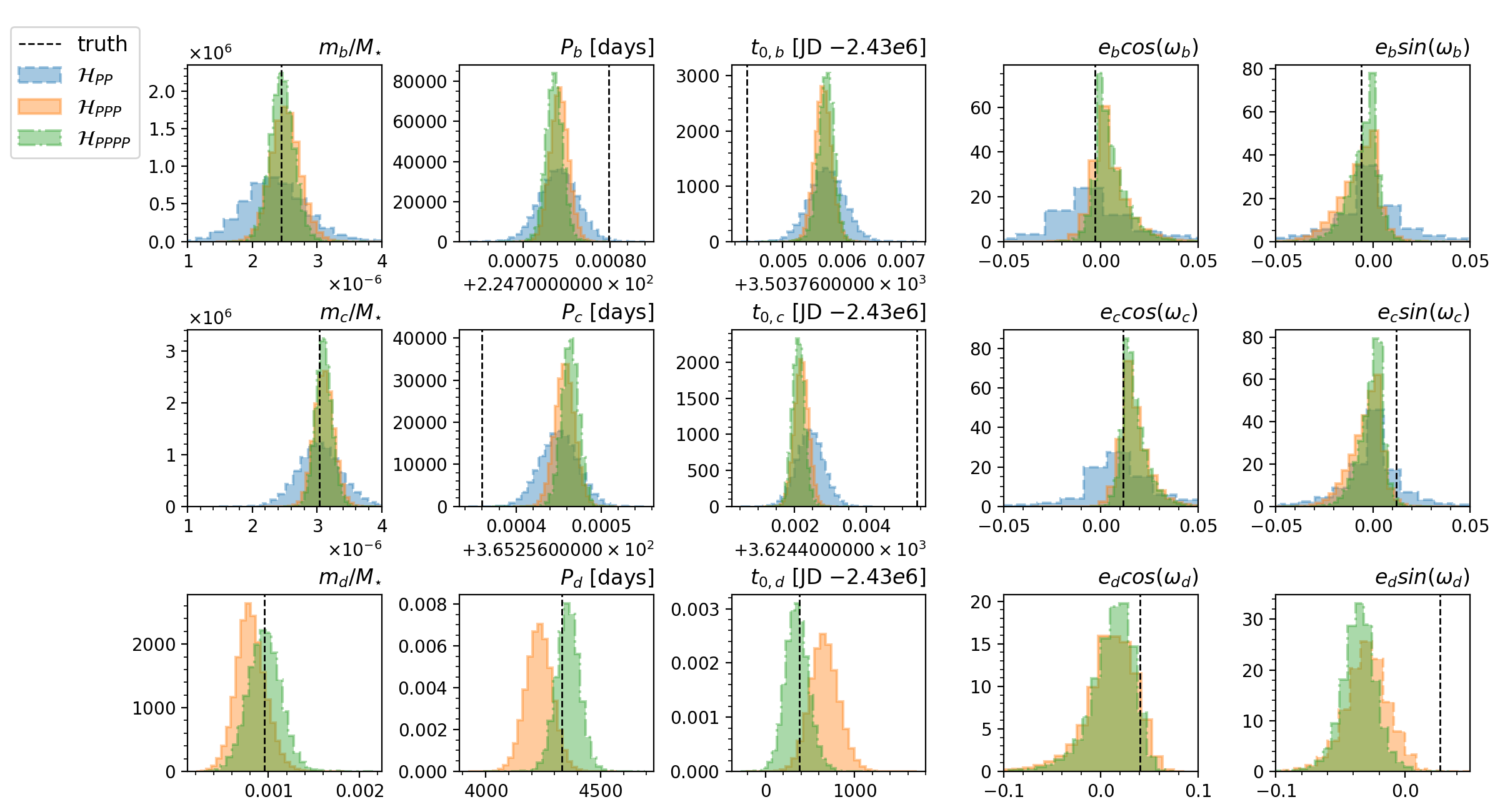}
    \caption{1D posterior histograms for planet parameters fit in at least two of our configurations for our fiducial case -- i.e. those common between our 2-, 3-, and 4-planet models discussed in \mbox{\S\ \ref{sec:ex_search}} -- compared to the truth (dashed vertical line) for Venus, Earth + Moon, and Jupiter. The rows correspond to planets b/c/d; the parameter being plotted is indicated in the title of each panel.} 
    \label{fig:1d_hist}
\end{figure*}

\begin{figure*}
    \centering
    \begin{tabular}{cc}
    \includegraphics[width=.5\linewidth]{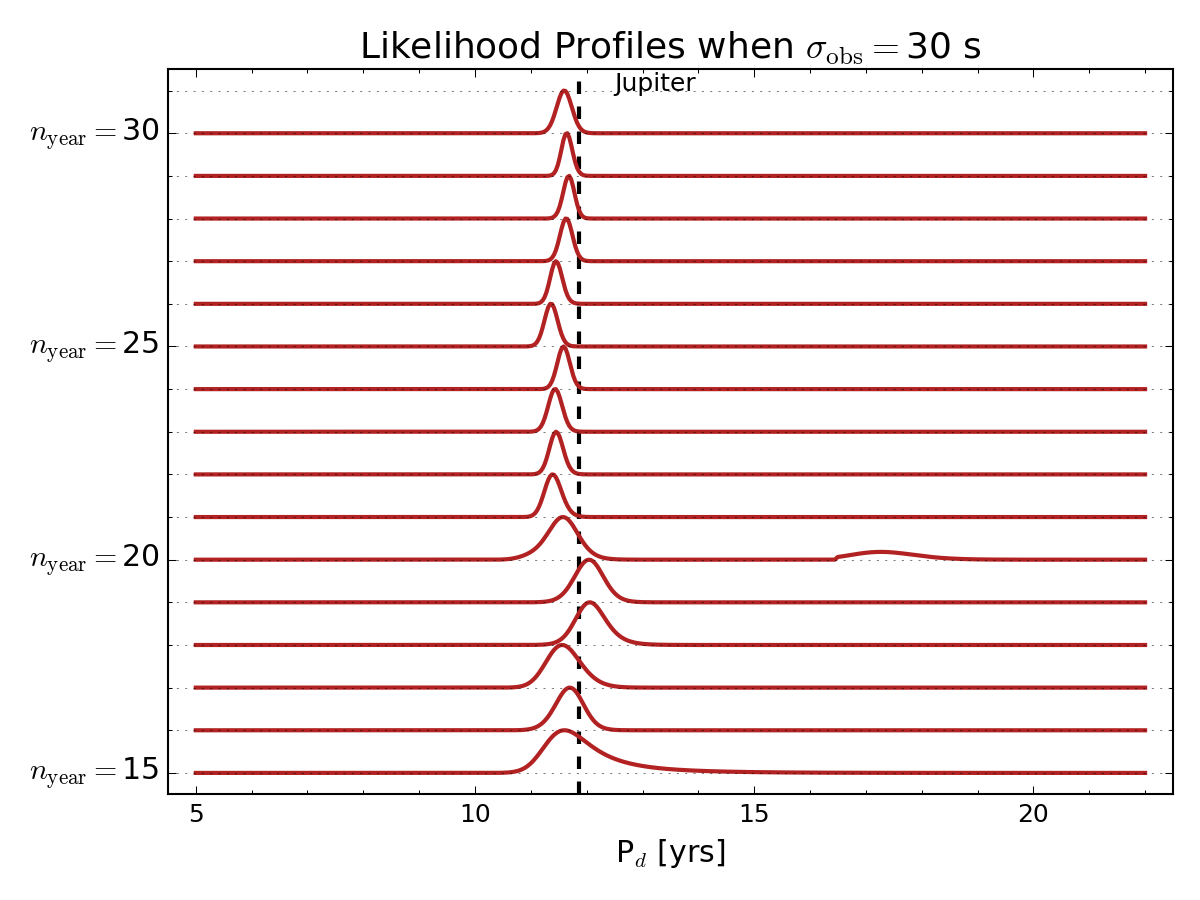} &
    \includegraphics[width=.5\linewidth]{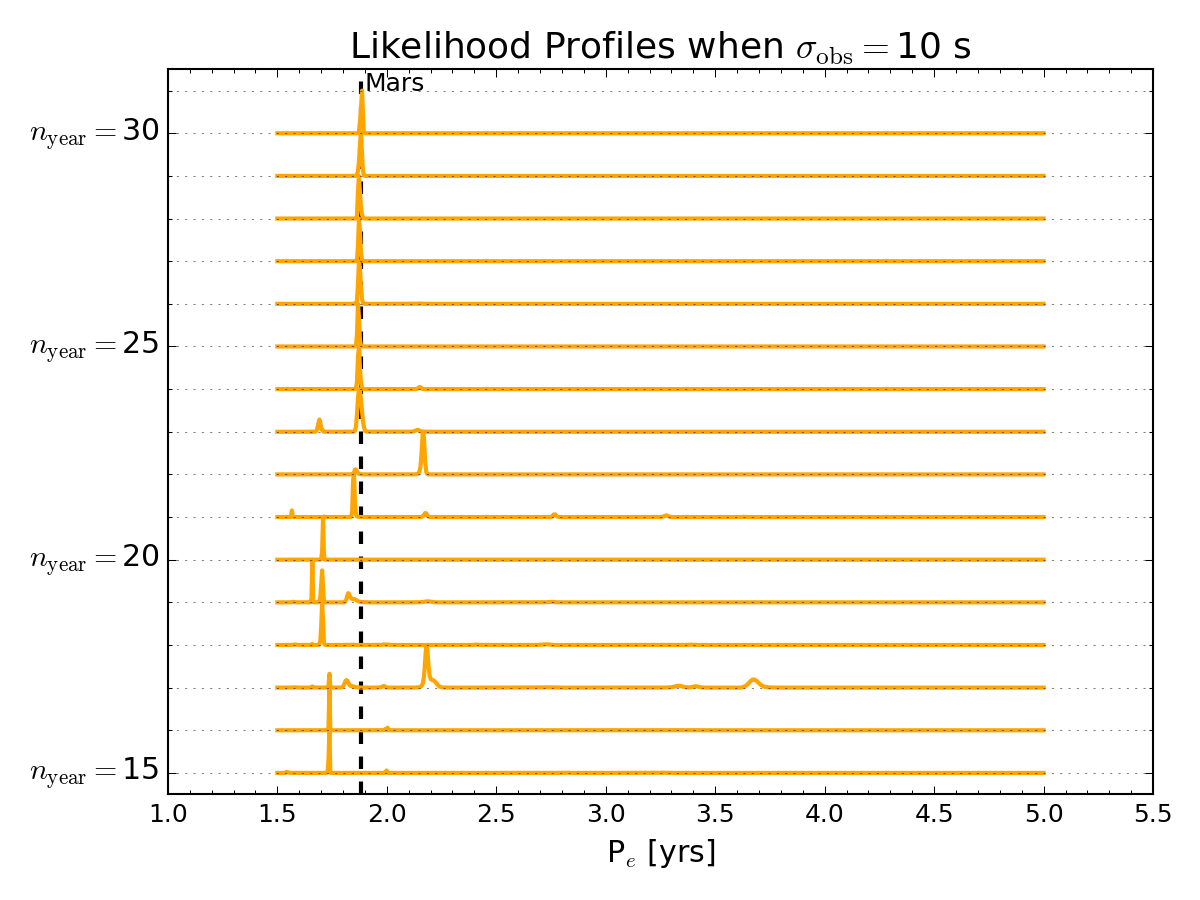} \\
    \end{tabular}
    \caption{Relative likelihood profiles for baselines simulated in this work, as indicated in the y-axis. \textit{Left}: the three planet search when  $\sigma_{\text{obs}} = 30$ seconds. \textit{Right} the four-planet search when $\sigma_{\text{obs}} = 10$ seconds.}
    \label{fig:multipeaks}
\end{figure*}

\begin{figure*}
    \centering
    \includegraphics[width=\hsize]{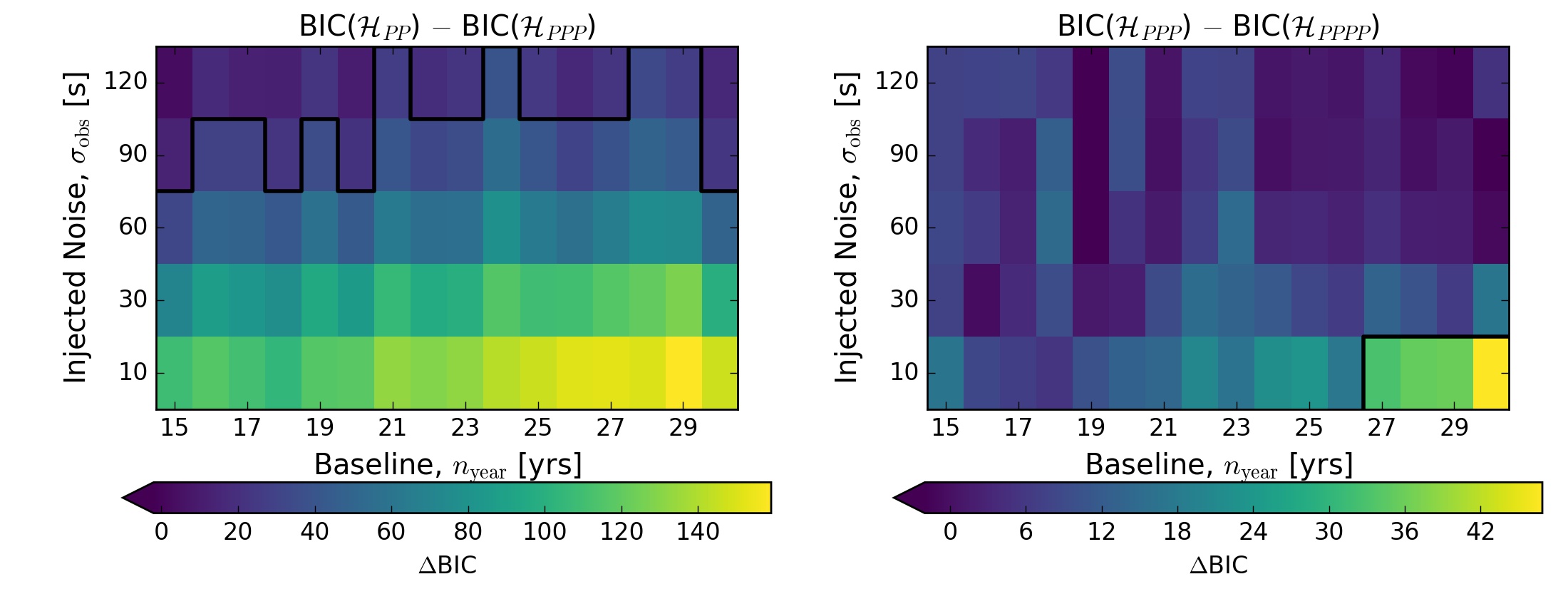}
    \caption{$\Delta$BICs for our best-fit models from the fine grid search (\S\ \ref{sec:tests}), where positive values represent a preference for an additional planet. Values below the black contour line correspond to at least $5\sigma$ detection of a third, massive planet (left) and a nearby terrestrial (right). Note the different scales of the colorbars.} 
    \label{fig:deltaBIC}
\end{figure*}

\begin{figure*}[ht]
\centering
    \begin{tabular}{cc}
    \includegraphics[width=.48\linewidth]{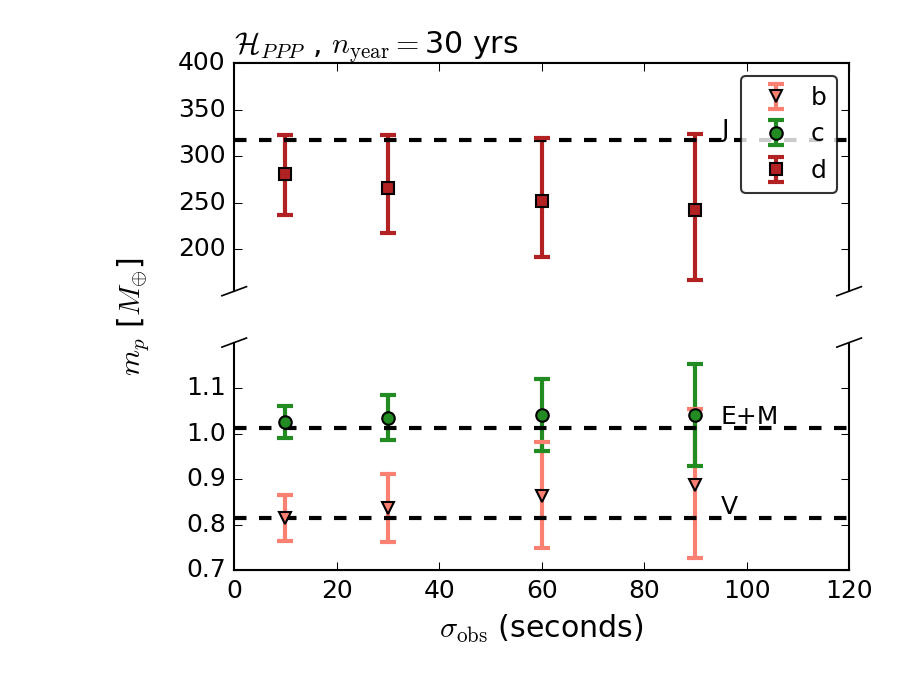} &  
    \includegraphics[width=.4\linewidth]{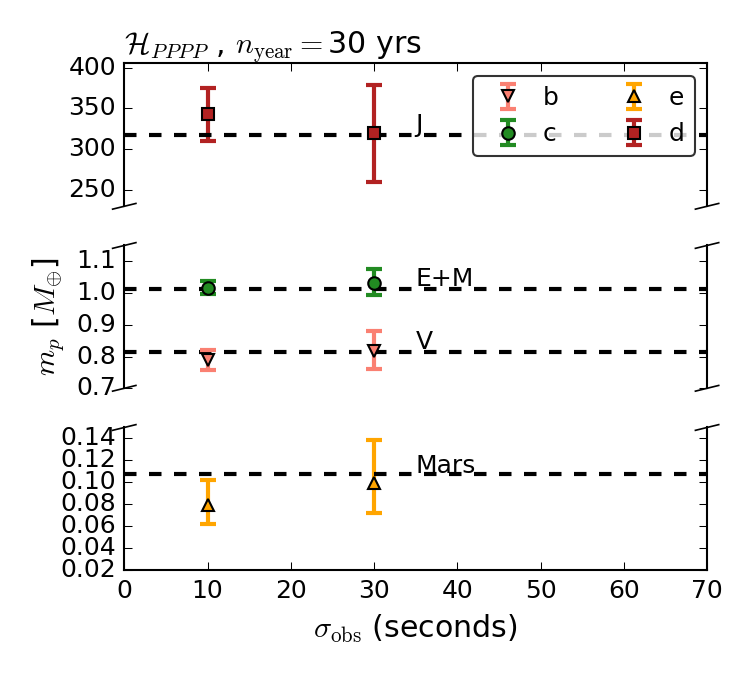} \\
    \end{tabular}
    \caption{Recovered median and $1\sigma$ posterior mass versus injected timing error for planets (b-e, respectively) around a Sun-like star: \hppp\ (left) and \hpppp\ (right) models. The true masses for Venus, Earth + Moon, Jupiter, and Mars are plotted as dashed horizontal lines. Note the different scales for the axes.} 
    \label{fig:masses_30}
\end{figure*}
\subsection{Masses and orbital elements}\label{sec:retrievals}
Generally, as more planets are included, the range of uncertainties on variables fit in the previous model is reduced. Figure~\ref{fig:1d_hist} contrasts the posterior distributions of the common parameters between our three hypotheses. The data used here represent our fiducial case of observations conducted over a 30-year baseline with 30-second timing errors. In the first and second row of panels -- corresponding to planets b and c -- we find that the posterior mass distributions for models \hppp\ and \hpppp\ are nearly identical, and narrower than those of the \hpp\ model. The true masses and Keplerian orbital elements for the relevant SS objects, provided in the last column of Table~\ref{tab:30s30yrs}, are plotted as dashed vertical lines. From these, we see that the planetary masses and eccentricities are very accurate for planets b/c: our Venus and EM analogues. 

\subsection{Effect of timing noise and observing baseline}\label{sec:effect}
The retrieval simulations in the previous subsection are all for 30 year baselines with 30 seconds of Gaussian white noise injected. Here we explore how the results vary across our simulations. 

For starters, our high probability region for the third planet's period are consistent for all baselines when $\sigma_\text{obs}=30$, albeit with wider distributions at shorter baselines, as shown in the left panel of Fig.~\ref{fig:multipeaks}. 

The difference in BICs suggest that \hppp\ is always preferred over the model with only two planets. In Figure~\ref{fig:deltaBIC}, bins below the black contour line correspond to a $5\sigma$ detection of planet d (left panel) and planet e (right panel). While there is evidence for a fourth planet, a statistically convincing detection is only possible for the unrealistically high timing precision lower than 20 seconds. At this noise level, the highest probability period for the fourth planet does not agree with that of Mars when $n_{\text{year}}\leq22$ (Fig.~\ref{fig:multipeaks}). We note that $\Delta$BIC values that suggest a preference for 4 planets when the timing errors are high ($\geq 60$) do not refer to models that retrieve Mars. Instead, these optimized solutions correspond to highly eccentric, massive planets.

We plot how the white noise affects our posterior masses, given a 30-year observing duration, in Figure~\ref{fig:masses_30}. For the \hppp\ points on the left, the systematic timing errors are 27.4, 30.2, 32.4, and 35.1 seconds; the systematic errors for \hpppp\ are 16.5, and 15.7 seconds (right panel). As expected, an increase in injected noise leads to larger mass uncertainties for a given model. Also, the measured masses for all the planets remain consistent between \hppp\ and \hpppp\ for all the noise levels. We do not plot retrieved masses for timing errors larger than 30 seconds for the \hpppp\ model because the Markov chains did not converge for those simulations.       

\begin{figure*}
    \centering
    \includegraphics[width=.8\hsize]{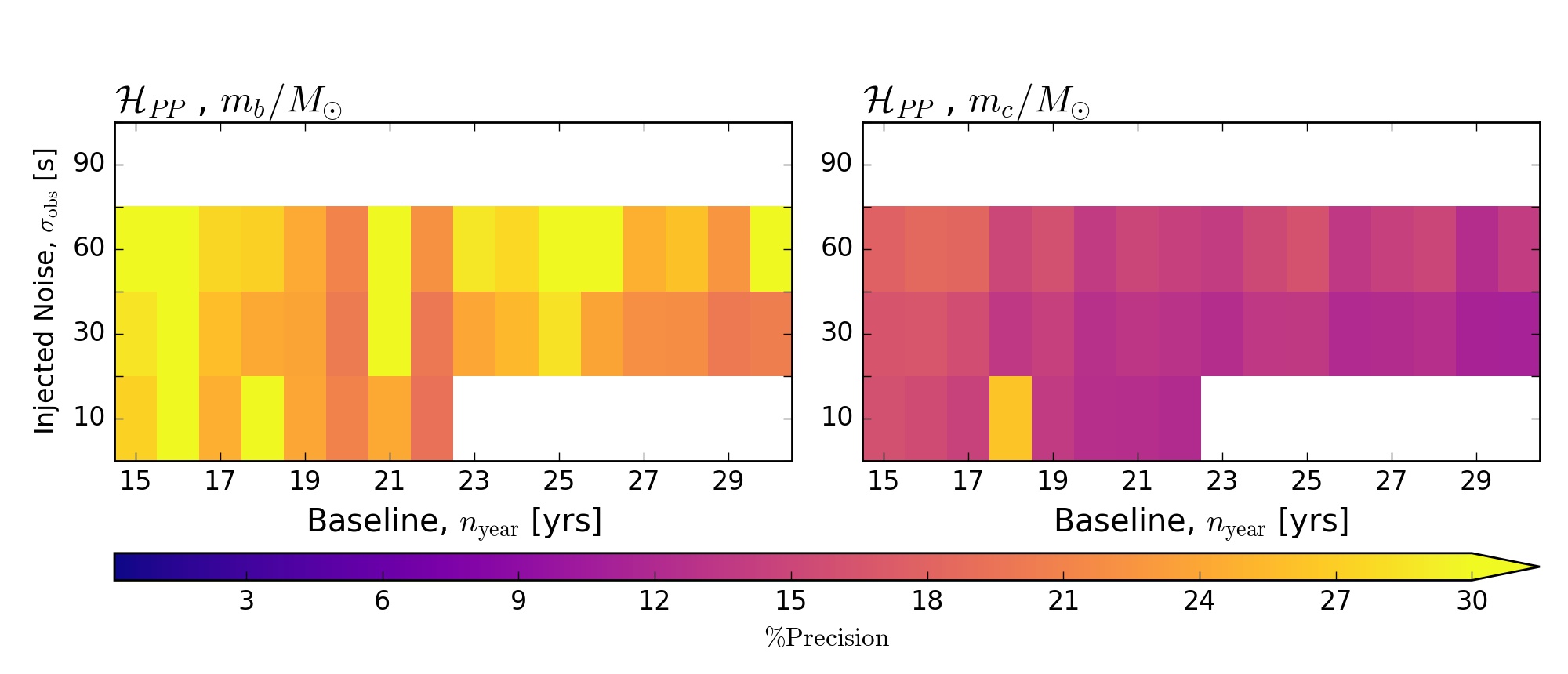}
    \caption{Percent precision on mean mass-ratio for planets b (left panel), and c (right panel) in the two-planet model, binned by white noise and observing baseline. Empty bins indicate where the chains did not find a solution, and we use the same colorbar scaling for both subfigures.}
    \label{fig:composite_p2}
\end{figure*}

\begin{figure*}
    \centering
    \includegraphics[width=\hsize]{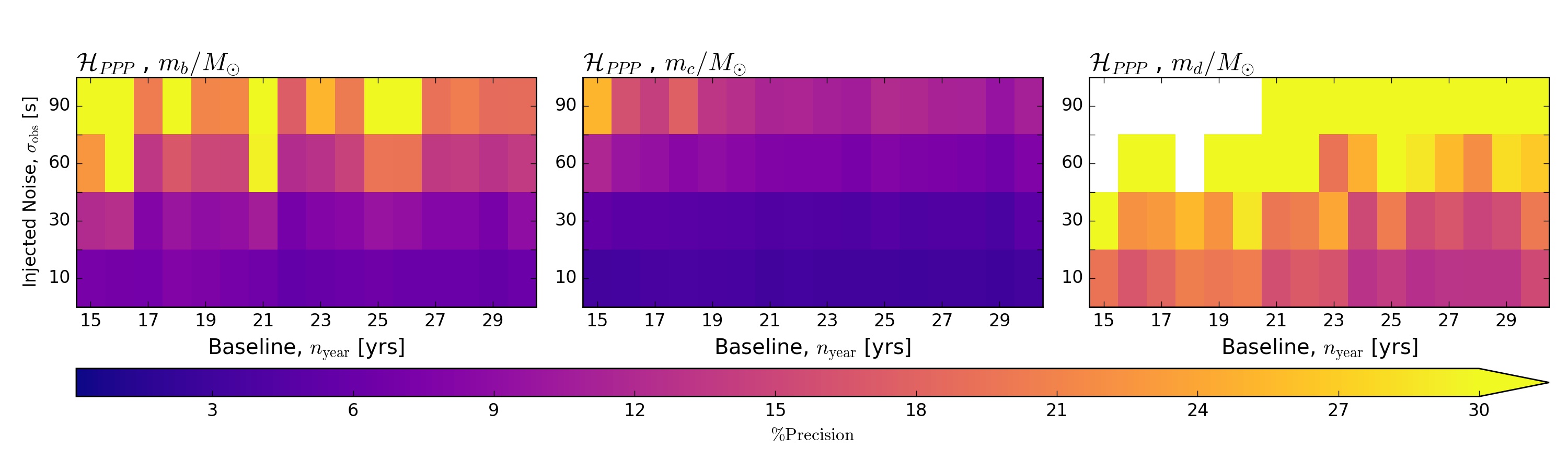}
    \caption{Same as in Fig.~\ref{fig:composite_p2}, but here we plot the percent precision of the best 3-planet model: planet b on the left), c in the middle, and d on the right. After fitting our fiducial simulation -- transits with 30-second injected noise observed over 30 years -- with a third planet, there is a reduction from 20\% to 8\% error for c, and from 11\% to 4\% error for b. Here, empty bins show when the percent precision was greater than 100\%.} 
    \label{fig:composite_p3}
\end{figure*}

More generally, figures \ref{fig:composite_p2} and \ref{fig:composite_p3} show how the mass-ratios precision change for 2- and 3-planet fits to our simulation set. In these plots, we defined  $\% \mathrm{Precision} = (\sigma_\mu / \overline{\mu})\times 100\%$, -- where we have measured the mean, $\overline{\mu}$, and standard deviation, $\sigma_\mu$ of the posterior mass-ratio distribution -- and use the same colorbar scale for each panel. For the same simulations (i.e. same realizations of noise), the addition of the third planet to the model greatly improves the precision for planets b/c.  That planet b's posteriors have higher uncertainties than planet c is likely due to fewer measured transits of Earth.

In contrast to the transiting exoplanets, the gas giant's parameters appear much less accurate and precise in the 3-planet model. Upon visual examination of the traces -- and employing the Gelman, Rubin, and Brooks diagnostic \citep{BrooksGelman1998} -- we found that the Markov chains did not actually converge for survey baselines shorter than 22 years (for 60 seconds of injected noise) and 23 years (when $\sigma_{obs}=90$). Therefore, we could not directly compare the percent errors for all the datasets. 

\begin{figure*}
    \centering
    \includegraphics[width=\hsize]{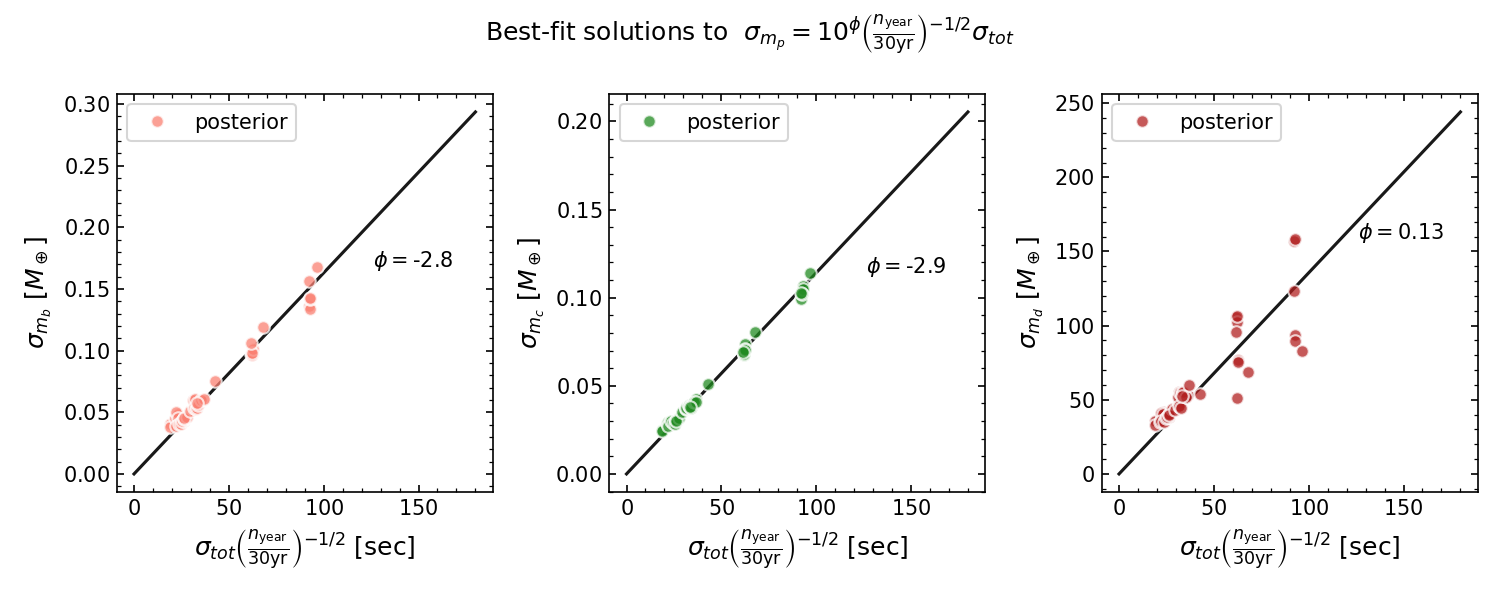}
    \caption{Power law relations for $\sigma_{m_p}$-$\sigma_{\text{tot}}$ derived from the posteriors of the 3-planet model: planet b (left), c (middle), and d (right). 
     For the x-axis, the injected timing noise has been added in quadrature to the mean of the posterior systematic error to obtain the total timing precision.}
    \label{fig:interp_errors}
\end{figure*}

For the converged chains of ${\cal H}_{PPP}$, we fit the following power law: 
\begin{equation}\label{equ:mass_corr}
\frac{\sigma_{m_p}}{m_\oplus} = 10^{\phi} \bigg(\frac{n_{\text{year}}}{30}\bigg)^{-1/2}  \bigg(\frac{\sigma_{\text{tot}}}{\mathrm{sec}}\bigg) \text{,}
\end{equation} 
where $\sigma_{m_p}$ is the standard deviation of mass, and the total timing precision is equal to $\sqrt{\sigma_\text{obs}^2+\overline{\sigma}_\text{sys}^2}$. The solutions to this equation allow us to predict the mass precision for a given observing baseline ($\ge$ 22 years) and timing precision. We plot these relations for each of the three planets in Figure~\ref{fig:interp_errors}. For the terrestrials, $\phi=-2.94$ and $\phi=-2.79$, respectively. We note that Equation~\ref{equ:mass_corr} assumes the Sun's mass is precisely known.

\begin{figure}
    \includegraphics[width=1.05\hsize]{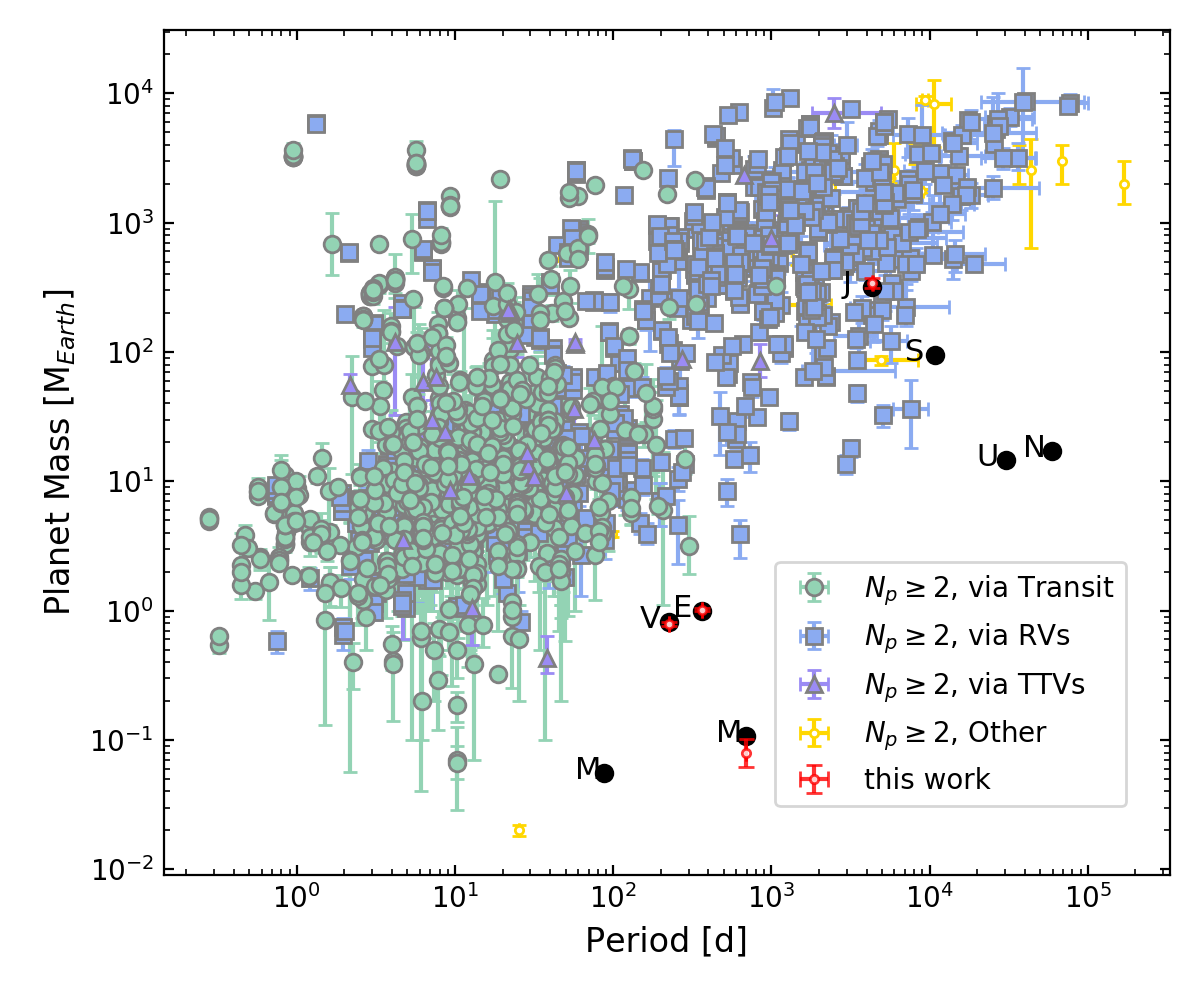}
    \caption{Measured planet mass as a function of orbital period for multi-planetary systems with $N_p$ planets, listed by discovery method: transit (green circles), RVs (blue squares), TTVs (purple triangles), and others (yellow dots) \citep{exo_archive}. The retrieved mass measurements in this work -- for Venus, EM, Mars, and Jupiter -- are taken from the right panel of Figure~\ref{fig:masses_30}: the 4-planet model with a 10-sec noise injected (i.e. a total timing uncertainty of 19.3 seconds.) Data accessed on 2024 Dec 29.}
    \label{fig:systems}
\end{figure} 

\section{Discussion}\label{sec:discussion}
\subsection{Model Accuracy }\label{sec:consistency} 
In this work, we find that the retrieved posteriors of the transiting planets are similar across the test configurations. In addition, the precision of the masses should improve as the timing errors become more precise, and as the model becomes more accurate, with the inclusion of more planets. 

Specifically, for Venus and EM, the measured masses are consistent with the correct values across the models with 2-4 planets. Since the 3- and 4-planet models better reproduce the actual TTVs, the systematic errors become smaller, hence improving the precision of their masses. The same applies to these planets' eccentricities, which become more precise and accurate as the model is improved from 2 to 3 to 4 planets. Thus, we expect that a long-term campaign to measure the masses of the pair of transiting planets is robust to the timing uncertainty and the number of planets in the model; that is, the inferred masses should be accurate.

As far as Jupiter is concerned, our results are mixed. In the 3-planet model, we detect a gas giant at high significance for many baseline-noise combinations. However, the mass of Jupiter is not retrieved precisely (i.e. $\sigma_{m_d} =\pm 70 M_\oplus$) until either total timing precision better than 50 seconds is achieved, or a 4-planet model is used. 
Despite its small mass, the presence of Mars does have an effect on the inference of the correct parameters of Jupiter; Mars is small, but it is (relatively) mighty.  This can be attributed to the closer proximity of Mars to the Earth-Moon-Barycenter, as well as its proximity to a 2:1 resonance, both of which enhance its dynamical influence on Earth relative that of Jupiter, whose long orbital period greatly diminishes its dynamical impact on Earth. Nevertheless, Mars is 3.5 orders of magnitude smaller in mass than Jupiter, so after Venus, Jupiter still dominates the TTVs of planet c. Thus, the detection of a giant planet with TTVs is more straightforward than its characterization.  

To accurately retrieve Mars, our simulation set requires an unrealistically optimistic level of precision for TTV measurement: equivalent to a 20-second transit-timing error. As such, we are pessimistic about the correct characterization of exoplanetary Mars-analogues around Sun-like stars, perhaps until multi-wavelength observations at very high signal-to-noise are performed. 
\subsection{Implications for observational capabilities}\label{sec:impact}
To find systems with architectures analogous to our solar system, we need to improve the transit timing sensitivity of a telescope to better than 20 seconds for Earth/Venus/Sun-analogue transits with terrestrial companions. If we could achieve this, we could probe a region of exoplanet discovery space which is unreachable with past and current technology. While future coronagraphic space telescopes could detect solar-system analogues \citep{Dalcanton2015}, they would not necessarily be able to measure the sizes and masses of the planets. Furthermore, there are no existing instruments that can detect and continuously measure TTVs over the baselines required to precisely characterize a SS-analogue. 

Future telescopes may have the capability to detect Earth/Venus- analogue transits. For instance, the \textit{PLATO} mission will continuously monitor long-duration observation fields for up to three years -- for which it has been estimated that about a dozen Earth-sized planets with orbits between 250-500d may be found \citep{Heller2022,Matuszewski2023}. If each of these Earth-sized planets were accompanied by an interior Venus-analogue, then at least one Venus-Earth analogue could be detected. However, these analogues are unlikely to be well characterized. Assuming that transit timing uncertainty scales as $\sim 1/$Signal-Noise-Ratio \citep[SNR;][]{Holczer2016}, \textit{PLATO} is unlikely to reach $\mathrm{SNR}=10$ when integrating 3 transits of an Earth-like planet. If made, these detections will also need to be followed up with telescope(s) of much larger aperture and multi-wavelength capability to correct for stellar variability and obtain high-precision transit times over an extended duration. Such a monumental task is beyond what is currently achievable for a system which hosts a cool gas giant as well as multiple terrestrials within, or near, the HZ.

Realistically, one could measure the RV signals caused by a long-period giant planet which does not transit, if RV data are sufficiently precise \citep{Pepe2008}. This would complement and confirm a transit-timing analysis, and vice versa. Moreover, a joint fit with simulated RV monitoring of the Sun could improve mass constraints for the outer gas giant, and may reduce the observing baseline required to accurately measure the masses of Venus and EM. 

Figure~\ref{fig:systems} shows the measured masses for known planets by discovery method\footnote{NASA Exoplanet Archive, \url{http://exoplanetarchive.ipac.caltech.edu}}. We include masses inferred from the \hpppp\ model using data that was simulated over 30 years -- with 19.3-second total transit-timing error. For context, 200 RV observations over several years are needed to measure the semi-amplitude (and therefore the mass) of a transiting Venus in a Kepler-like system with better than 20\% precision using a next-generation RV instrument \citep[$10$ cm/sec precision;][]{He2021}. In Fig.~\ref{fig:systems}, the inferred 0.02 $M_\oplus$ mass-uncertainty for Mars -- which would require a radial-velocity precision of 1.45 mm/sec -- is beyond any planned capability of extreme precision radial velocity experiments. With the caveat that stellar variability of solar-like stars can swamp a planet signal, ESPRESSO can just detect Earth mass habitable-zone planets orbiting a 0.8\msun\ dwarf with a 10 cm/s RV precision, but not Mars-mass \citep[see][]{Pepe2013,VanEylen2021}. Thus, it remains to be seen whether future instruments can characterize the masses of a system with both habitable-zone terrestrial planets and gas giant planets around a G dwarf, like our solar system. 

\subsection{Future work}\label{sec:future}
In its current state, \ttvfaster\ breaks down in accuracy at high eccentricities, for large planet-star mass ratios, and for planets close to resonance \citep{Agol2016}. Therefore, future work could involve implementing a more accurate model, such as the $\texttt{NBodyGradient}$ code \citep[see][]{Agol2021} to analyze the simulated solar system. Similarly, a full photo-dynamical simulation of realistic transits with varying levels of white noise and correlated noise appropriate for our Sun could help us determine how precisely the times can be measured \citep[a la][but with bigger glass or multiple wavelengths to achieve higher precision]{Morris2020}. We leave investigating the impact of missed transits and transit-timing outliers to future work.

\section{Conclusion}\label{sec:conclusion}
To date we have yet to detect and characterize an exoplanet system that is analogous to our solar system.  In principle, it should be possible to detect the transits of both Earth and Venus as seen from afar; and so here we have investigated what might be learned about their masses and the architecture of the solar system from long-term, high-precision transit-timing measurements of Venus and Earth+Moon (we leave exploration of the Moon's influence on Earth to future work). The AD16 model is co-planar, which is suitable for our analysis because mutual inclinations and eccentricities of the relevant solar system bodies are small. This work could possibly help motivate mission development, such as the \textit{Nautilus} concept \citep{Apai2019}.

Our focus was to compute the simultaneous detectability of an additional non-transiting gas giant and terrestrial planet, given detectable TTV signals from two transiting rocky planets. Our computed transit times are fully available online, along with the code used to analyze them\footnote{\dataset[DOI: 10.5281/zenodo.17156392]{https://doi.org/10.5281/zenodo.17156392}; \url{https://github.com/bmlindor/ttv_ss}}. The results are as follows: 
\begin{itemize}
    \item We recover the correct masses and orbit shapes for Venus and EM with less than 50 transit total measurements. Their measured masses appear to be robust; that is, their masses are accurate whether or not we include additional planets in the transit-timing model. However, their masses become more precise when we include Jupiter in the transit-timing model. Therefore, perturbations by an unseen, distant gas giant can lead to an additional source of uncertainty in existing transit-timing models.
    \item We can readily detect a wide-separation gas giant with transit observations spanning only 1.25 orbits of the giant planet with $\sim 90$-second noise. However, measuring a reliable mass  ($\pm 70 M_\oplus$) for a Jupiter-analogue requires better than 50-second total error, and/or including Mars in our transit-timing model.
    \item We would require better than 20 second timing precision to detect and correctly characterize a system analogous to ours -- but only for the planets out to several astronomical units, excluding Mercury.
    \item We provide equations to estimate the mass precision of Venus-, Earth- and Jupiter-like planets in a system like ours, given a transit timing precision and an planned observing baseline. At the 86-second stellar variability noise floor predicted for \textit{PLATO} \citep{Morris2020}, we could measure $1\sigma$ mass uncertainties of 0.24 and 0.19 $M_{\oplus}$ for 30-year long observations of a transiting Venus and Earth -- which is only possible if a Jupiter-analogue is included in the model.
\end{itemize}

\section*{Acknowledgements}
This material is based upon work supported by the NSF Graduate Research Fellowship Program under Grant No.\ DGE-1762114.
E.A.\ acknowledges support from NSF grant AST-1907342, NASA NExSS grant No.\ 80NSSC18K0829, and NASA XRP grant 80NSSC21K1111. This work was performed in part by the Virtual Planetary Laboratory Team, which is a member of the NASA Nexus for Exoplanet System Science, and funded via the NASA Astrobiology Program ICAR Grant 80NSSC23K1398.
This research has made use of the NASA Exoplanet Archive, which is operated by the California Institute of Technology, under contract with the National Aeronautics and Space Administration under the Exoplanet Exploration Program.

We thank Steven G. Johnson for creating packages to access Python functions and modules from the Julia language.
We thank Matthew Holman, Stephen Kane, and Brett Morris for useful discussions. 
We also thank the reviewers for their thoughtful comments which helped to improve this manuscript. 

\software{
\calceph\, \citep{Gastineau2015:CALCEPH},
\ttvfaster\ \citep{Agol2016}, 
\texttt{matplotlib} \citep{Hunter2007,Caswell2019},
Julia \citep{Bezanson2017},
\texttt{DataFrames.jl}\citep{DataFrames}
}
\bibliographystyle{psj}
\bibliography{refs.bib}

\appendix
\section{Posteriors}
\begin{figure*}
\center
\includegraphics[width=\hsize]{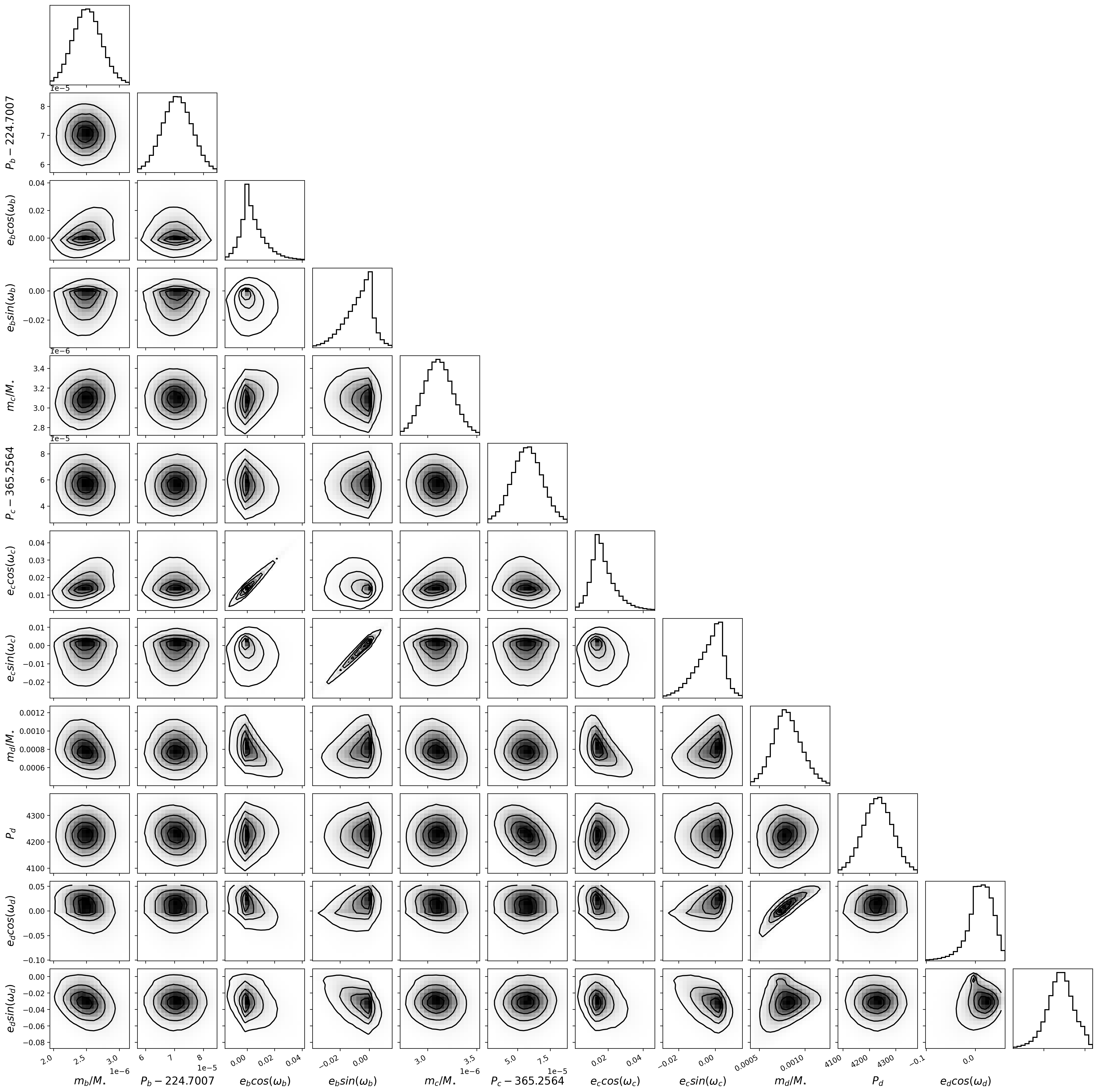}
\caption{Posterior probability density functions of the \hppp\ model, which includes the mass-ratios, periods, and eccentricity-vectors of each planet. In the panels along the diagonal, we bound the histograms by the 99.5\% confidence interval of a given component. The off-diagonal panels show parameter correlations, and the 0.5-,1-,1.5-, and 2-$\sigma$ contours.}
\label{fig:cornerplot_p3}
\end{figure*}

\begin{figure*}
\center
\includegraphics[width=\hsize]{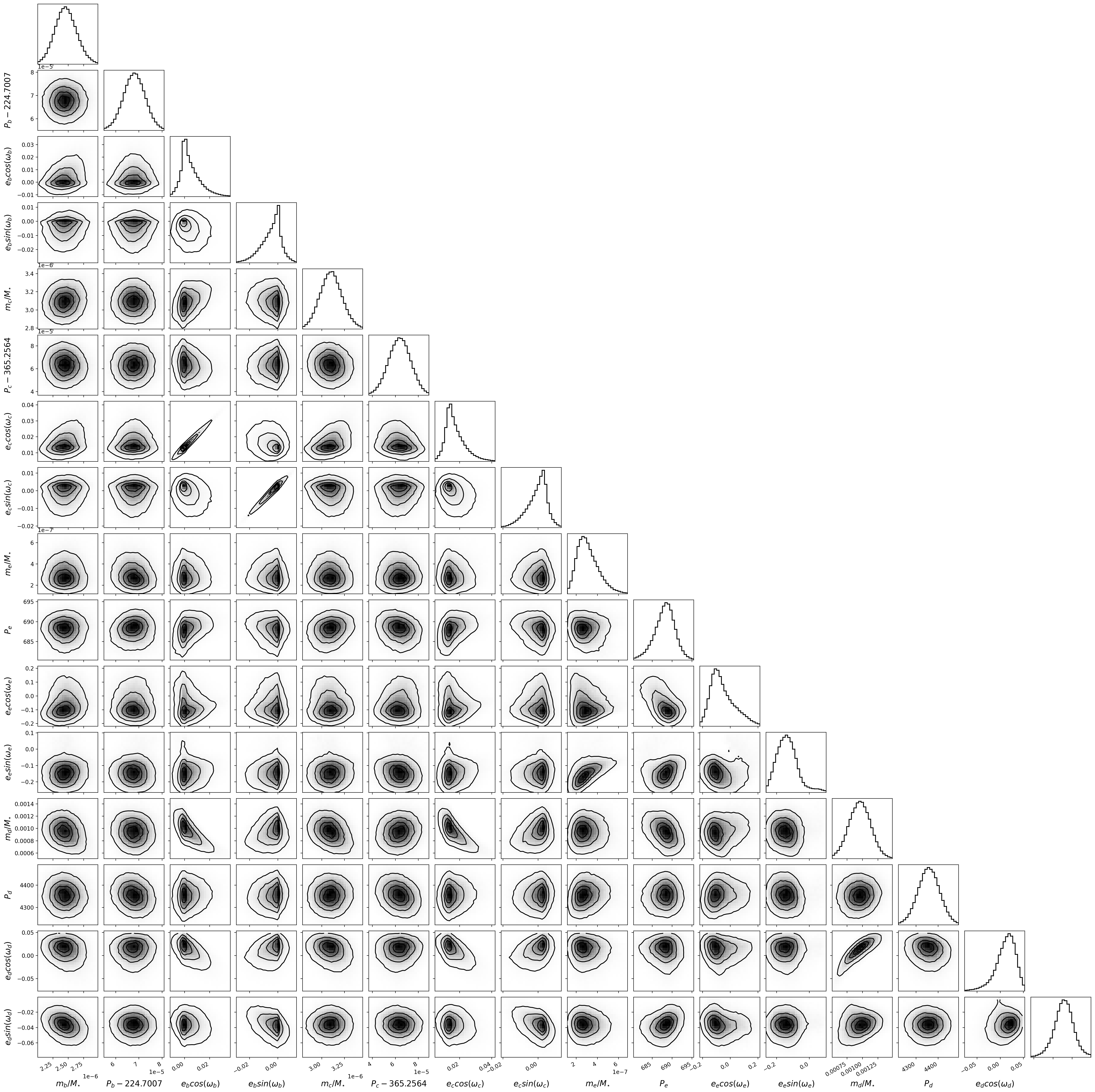}
\caption{Same notes as Fig~\ref{fig:cornerplot_p3}, but this corner plot corresponds to the \hpppp\ model.}
\label{fig:cornerplot_p4}
\end{figure*}

\begin{table*}[!ht]
\centering
\caption{Median and $1\sigma$ parameters, listed by test configuration. We also report the ${\chi_0}^2$ at the model parameters with the maximum likelihood; time of periastron passage ($t_0$) is expressed as $\text{JD} - 2430000$.}
\label{tab:30s30yrs}
\begin{threeparttable}
\begin{tabular}{cccccc}
\hline
\hline
Parameter & \hpp\ & \hppp\ & \hpppp\ & Truth\\ \\ 
\hline 
$\chisq_0$ & 582.9 & 109.6 & 77.39 & \\ BIC & 532.3 & 432.8 & 415.8 \\ \hline
\multicolumn{1}{l}{Fitted Parameters}  & & & & \\
$\sigma_\text{sys}$ [sec]	& $85.8_{-7.51}^{+8.67}$ & 	$30.2_{-5.33}^{+5.65}$ & 	$15.7_{-8.11}^{+6.9}$ \\
\multicolumn{1}{l}{}  & & & & \textbf{Venus}\\$\mu_b \times 10^{-6}$	& $2.3375_{-0.4452}^{+0.4841} $ & 	$2.5111_{-0.2188}^{+0.2237} $ & 	$2.4586_{-0.1740}^{+0.1840} $ &	2.4464\\
$P_b$ [days]    &	$224.70077\pm{0.00001} $ & 	$224.700771\pm{0.000005}$ & 	$224.700780\pm{0.000005} $ &	224.7007\\
$t_{0,b}$ [days]    &	$3503.7657\pm{0.0003} $ & 	$3503.7656\pm{0.0001} $ & 	$3503.7657\pm{0.0001} $ &	3503.764419\\
$e_b \cos{\omega_b}$    &	$-0.0041_{-0.0174}^{+0.0211} $ & 	$0.0023_{-0.0058}^{+0.0104} $ & 	$0.0030_{-0.0047}^{+0.0096} $  & 	-0.003\tnote{\textsuperscript{*}}\\
$e_b \sin{\omega_b}$    &	$-0.0015_{-0.0158}^{+0.0126} $ & 	$-0.0056_{-0.0117}^{+0.0067} $ & 	$-0.0033_{-0.0087}^{+0.0047} $ & 	-0.006\tnote{\textsuperscript{*}}\\

\multicolumn{1}{l}{}  & & & & \textbf{Earth+Moon}\\
$\mu_c \times 10^{-6}$  &	$3.0385_{-0.3206}^{+0.3372} $ & 	$3.1081_{-0.1485}^{+0.1517} $ & 	$3.1034_{-0.1216}^{+0.1251} $ &	3.0369\\
$P_c$ [days]    &	$365.256450\pm{0.000002} $ & 	$365.25646\pm{0.00001} $ & 	$365.25646\pm{0.00001} $ &	365.25636\\
$t_{0,c}$ [days]    &	$3624.4024\pm{0.0004} $ & 	$3624.4022\pm{0.0002} $ & 	$3624.4021\pm{0.0002} $ &	3624.405369\\
$e_c \cos{\omega_c}$    &	$0.0096_{-0.0125}^{+0.0180} $ & 	$0.0159_{-0.00488}^{+0.0081} $ & 	$0.0158_{-0.00401}^{+0.0075} $ &	0.011\tnote{\textsuperscript{*}}\\
$e_c \sin{\omega_c}$    &	$0.0002_{-0.0121}^{+0.0103} $ & 	$-0.0019_{-0.0095}^{+0.0057} $ & 	$0.0004_{-0.0071}^{+0.0041} $ & 	0.012\tnote{\textsuperscript{*}}\\
\multicolumn{1}{l}{}  & & & & \textbf{Jupiter} \\$\mu_d$	 & -- & $0.000801_{-0.000148}^{+0.000169} $ & 	$0.0009592_{-0.000179}^{+0.000177} $  & 	0.00095\\
$P_d$ [days]	 & -- & $4232.5_{-56.999}^{+56.448} $ & 	$4358.274_{-49.333}^{+49.427} $ &	4332.82\\
$t_{0,d}$ [days]	 & -- & $658.369_{-154.062}^{+162.989} $ & 	$352.408_{-123.909}^{+131.818} $ & 	333.7268\\
$e_d \cos{\omega_d}$	 & -- & $0.0123_{-0.0251}^{+0.0227} $ & 	$0.0145_{-0.0233}^{+0.0177} $ & 	0.0403\tnote{\textsuperscript{*}}\\
$e_d \sin{\omega_d}$	 & -- & $-0.0294_{-0.0165}^{+0.0159} $ & 	$-0.0354_{-0.0124}^{+0.0118} $ & 	0.0268\tnote{\textsuperscript{*}}\\
\multicolumn{1}{l}{}  & & & & \textbf{Mars}\\$\mu_e \times 10^{6}$	& 		--&	--& $0.2983_{-0.0829}^{+0.1176}$ & 0.3227	\\
$P_e$ [days]		&	--& 	--&	$688.347_{-2.692}^{+2.302} $ & 686.980\\
$t_{0,e}$ [days]	&	--& 		--&	$383.548_{-30.090}^{+28.042} $ & 383.823\\
$e_e \cos{\omega_e}$	& 	--&	--&	$-0.0712_{-0.0645}^{+0.1138} $  & 0.0131\tnote{\textsuperscript{*}}\\
$e_e \sin{\omega_e}$	&		--& 	--&	$-0.1388_{-0.0587}^{+0.0621} $ & 0.0925\tnote{\textsuperscript{*}}\\
\hline
\textit{Derived Parameters}  & & \\
$m_b [\mearth]$ & $0.778^{+0.161}_{-0.148}$ & $0.836^{+0.074}_{-0.073}$ & $0.819^{+0.061}_{-0.058}$ & 0.815\\
$m_c [\mearth]$ & $1.012^{+0.123}_{-0.107}$ & $1.035^{+0.051}_{-0.049}$ & $1.033^{+0.042}_{-0.040}$ & 1.012\\
$m_d [\mearth]$ & -- & $266.532^{+56.084}_{-49.141}$ & $319.337^{+58.878}_{-59.440}$ & 317.8\\
$m_e [\mearth]$ & -- & -- & $0.099^{+0.039}_{-0.028}$ & 0.107 \\

$e_b$ & $0.019_{-0.002}^{+0.061}$ & $0.011_{-0.001}^{+0.020}$ & $0.008_{-0.001}^{+0.018}$ & 0.006 \\
$e_c$ & $0.016_{-0.001}^{+0.051}$ & $0.018_{-0.003}^{+0.013}$ &  $0.017_{-0.003}^{+0.012}$ & 0.016\\
$e_d$ & -- &  $0.040_{-0.008}^{+0.025}$ & $0.043_{-0.035}^{+0.016}$ & $0.048$\\
$e_e$ & -- & -- & $0.178_{-0.035}^{+0.057}$ & 0.09\\
\hline
\hline
\end{tabular}
\begin{tablenotes} 
\item[\textsuperscript{*}] The true $\omega$ components of the eccentricity vectors at J2000 have been corrected to match the reference frame of the JPL ephemerides by adding 77$^\circ$.
\end{tablenotes}
\end{threeparttable}
\end{table*}
We provide the MCMC parameter median and $1\sigma$ confidence intervals from fitting the transit times of Venus/Earth across the Sun over 30 years. This dataset had $\sigma_\mathrm{obs} = 30$ second Gaussian noise injected (discussed in \S \ref{sec:ex_search}). Figures \ref{fig:cornerplot_p3} and \ref{fig:cornerplot_p4} show corner plots of parameters in the 3-planet, and 4-planet models reported in Table~\ref{tab:30s30yrs}.  In the table's right-most column, we provide masses and Keplerian orbital elements for Venus, Earth + Moon, Jupiter and Mars -- adopted  from \citet{Hussmann2009} or calculated using JPL ephemerides \citep{Park2021}.

\end{document}